\documentclass[amsmath,11pt]{article}
\usepackage{amsfonts,amsmath,amssymb,graphics,epsfig}
\usepackage{enumerate}\usepackage{theorem}

\date{\empty}

\begin{document}

\title{\bf How the magnetic field behaves during the motion of a highly conducting fluid under its own gravity--A new theoretical, relativistic approach}

\author{Panagiotis Mavrogiannis${}^1$ and Christos G. Tsagas${}^{1,2}$\\ {\small ${}^1$Section of Astrophysics, Astronomy and Mechanics, Department of Physics}\\ {\small Aristotle University of Thessaloniki, Thessaloniki 54124, Greece}\\ {\small ${}^2$Clare Hall, University of Cambridge, Herschel Road, Cambridge CB3 9AL, UK}}

\maketitle

\begin{abstract}
  Within the context of general relativity we study in a fully covariant way the so-called Euler-Maxwell system of equations. In particular, on decomposing the aforementioned system into its 1 temporal and 1 + 2 spatial components at the ideal magnetohydrodynamic limit, we bring it in a simplified form that favors physical insight to the problem of a self-gravitating, magnetized fluid. Of central interest is the decomposition of Faraday's equation which leads to a new general solution governing the evolution of the magnetic field during the motion of the highly conducting fluid. According to the latter relation, the magnetic field generally grows or decays in proportion to the inverse cube law of the scale factor--associated with the continuous contraction or expansion of the fluid's volume respectively. The magnetic field's law of variation, which has remarkable implications for the motion of the whole fluid, is subsequently applied to homogeneous (anisotropic-magnetized) cosmological models--especially to the Bianchi I case--as well as to the study of homogeneous and anisotropic gravitational collapse in a magnetized environment. Concerning the cosmological application, we derive the evolution equations of Bianchi I spacetime permeated by large-scale magnetic fields (these equations reduce to their FRW counterparts at the small/large--scale limit). Also, the compatibility of the new evolution formula for the magnetic field with the standard cosmic nucleosynthesis constraint is examined. As for the application in astrophysics, our results predict that homogeneous gravitational implosion is impeded when the electric Weyl tensor (associated with tidal forces) along the magnetic forcelines overwhelms the magnetic energy density. Lastly, our model denotes that the satisfaction of the aforementioned criterion is ultimately driven into a problem of initial conditions.
\end{abstract}

\section{Introduction}

The question which triggered the present piece of work, though not directly related to the major part of its content, was whether the magnetised environment of a compact stellar object or of a protogalactic cloud could favor the inhibition of its gravitational collapse. The role which the magnetic fields play in such problems, is generally known in astrophysics. From the relativistic point of view however, it may be less known that magnetic fields acquire particular interest due to their direct coupling, as vectors, with the spacetime curvature~\cite{T1}--\cite{MavT}.

Previous independent relativistic studies have supported the following basic ideas regarding the behavior of magnetic fields in curved spacetimes. First, magnetic fields have the impressive ability not to self-gravitate; in other words, not to contract or collapse under their own gravity independently of the latter's strength~\cite{Mel, Th}. Second, in the presence of an external gravitational field, magnetic forcelines tend to stabilise themselves by developing naturally curvature related stresses which resist their gravitational deformation~\cite{Mel, Th}. Third, the key factor giving rise to such an unconventional behavior in both cases is the magnetic field's tension coming from the elasticity of its forcelines~\cite{T3, KT, TMav}.

Given the wide presence of magnetised fluids not only in the field of astrophysics but in cosmology as well, a primary question comes to the surface throughout all this past work. How does the magnetic field of a highly conducting fluid behave quantitatively or change due to the fluid's self-gravitating motion? Furthermore, if knowing its behavior, could we use it to extract information regarding the whole system-fluid and, subsequently to address realistic problems such as magnetised cosmological models and gravitational (astrophysical) collapse of charged matter\footnote{Besides, studies of collapsing charged matter have suggested that repulsive Coulomb forces could cause a bounce of the fluid, change of its contraction to an expansion, preventing thus the formation of singularities~\cite{N}-\cite{R}.}? This is basically the object of the present study.

Our proposed (covariant) approach to the problem consists of dealing with the Euler--Maxwell system of equations describing the motion of a magnetised fluid--at the ideal magnetohydrodynamic limit (for a tetrad-based approach to the problem, however with by far different aims, methodology and results see for instance~\cite{DK}). More specifically we study the system by decomposing its individual equations in one temporal and one plus two spatial components (one specific spatial direction and a 2-dimensional surface orthogonal to it). The mathematical context of our method is known as \textit{1+1+2 covariant relativistic approach}~\cite{CB}. First of all, the covariant approach to relativity differs from the more familiar \textit{metric based approach} in that the evolution equations, as well as the relevant constraints satisfied by the individual components of all spacetime quantities, are derived from the Ricci and the Bianchi identities, instead of the metric. Therefore, due to their geometric generality, the covariant formulae can be readily adapted to a wider spectrum of applications. In the second place, as already mentioned, it allows for access to details of the problem in question via the decomposition of the various mathematical objects (vectors, tensors, equations etc.) in components.

Being interested in the evolution of the magnetic field and its implications for the motion of the whole fluid, it is sufficient for us to focus on the Euler--Maxwell system of equations (actually supplemented by the so-called \textit{Raychaudhuri equation}) instead of the full Einstein--Maxwell system. By referring to the latter we mean the system consisting of the conservation laws (these are the so-called continuity and Euler's equations), coming from Einstein's field equations, and obeyed during the motion of a charged fluid; the propagation equations and the constraints, coming from Maxwell equations, and satisfied by the electric and magnetic components of the Maxwell field; an equation of state for the fluid--since now we have mentioned the equations which compose the Euler--Maxwell system--; the propagation equations and the constraints, coming from the Ricci identities for a fundamental, timelike 4-velocity field, and satisfied by the individual fluid dynamic quantities; finally, the propagation equations and the constraints, coming from the Bianchi identities, and satisfied by the individual components of the Weyl (long-range) curvature tensor. In practice, on decomposing the Euler-Maxwell equations in their individual temporal and spatial components, and on considering the ideal magnetohydrodynamic limit, the system takes a significantly simple form, not directly coupled with the long-range curvature (Weyl) terms. Therefore, we can achieve a first description of the charged fluid's motion without taking into account the long-range gravity effects. However, the latter are taken directly into account, in particular the electric Weyl curvature tensor, when studying the gravitational collapse of a highly conducting fluid in section~\ref{sec:Grav-collapse}. 

The present manuscript starts with a general presentation of the covariant approach to relativity, initially of its 1+3 form and subsequently proceeds to its extended 1+1+2 form. The emphasis is put on studying the dynamics of matter and electromagnetic fields as well as of their coupling. Some new details-developments (not taken from the literature) concerning the 1+1+2 decomposition, make part of the Appendix and provide a crucial supplement to the main text. After the theoretical introduction, we proceed to the decomposition and the detailed study of the Euler--Maxwell system of equations. We derive the relation describing the general evolution of the magnetic field and discuss its implications for the motion of a highly conducting fluid. Subsequently, we apply the latter, in the first place to the problem of homogeneous, magnetised cosmological models (section~\ref{sec:Cosmol-magn-fields}). In detail, the evolution formula for the magnetic field with respect to the scale factor is derived and subsequently used to find the expansion/contraction formulae of the Bianchi I cosmological model. Emphasis is given on determining the epoch of equality between magnetic energy density and matter/radiation in the aforementioned model. The epoch in question turns out to significantly differ (temporally) from its Friedmann counterpart. In parallel, the compatibility of the magnetic density evolution with the cosmic nucleosynthesis constraint is examined at an initial stage. In the second place, the magnetic field evolution formula in combination with the Raychaudhuri equation are used to investigate the problem of homogeneous and magnetised gravitational collapse (section~\ref{sec:Grav-collapse}). Our study points out the crucial role played by the magnetic energy density and the electric Weyl curvature in establishing a criterion which determines the fate of the collapse. Subsequently, the aforementioned criterion is tested in the context of a perturbed Bianchi I model of magnetised gravitational contraction. The results reveal the ultimately crucial role of the initial conditions in determining the implosion's outcome.

\section{The 1+3 covariant relativistic formalism}

In the present section we outline the basic principles of the \textit{1+3 covariant approach} (refer to the extensive reviews of~\cite{EMM} and~\cite{TCM}), we introduce the kinematic quantities and subsequently provide the background for the description of a charged, conducting fluid. The covariant approach to relativity, as described in the following, differs from the more familiar \textit{metric based approach} in that the evolution equations, as well as the relevant constraints satisfied by the individual components of all spacetime quantities, are derived from the Bianchi and the Ricci identities, instead of the metric. Therefore, due to their geometric generality, the covariant formulae can be readily adapted to a wider spectrum of applications.

\subsection{Background}

In the context of the 1+3 covariant formalism the 4-D relativistic space-time decomposes into a temporal direction and a 3-D space orthogonal to it. This space-time split is achieved by introducing a family of (\textit{fundamental}) observers who follow time-like orbits along curves (the so-called \textit{worldlines}) with local coordinates $x^{a}=x^{a}(\tau)$ where $a=0,1,2,3$ and the parameter $\tau$ is the observer's \textit{proper time}. The tangent (time-like) vector to the worldlines, $u^{a}\equiv dx^{a}/d\tau$ (normalised so that $u^{a}u_{a}=-1$), is called the observer's 4-velocity and it defines a temporal direction. Now if $g_{ab}$ is the metric of the 4-D space-time, a symmetric tensor can be defined, $h_{ab}\equiv g_{ab}+u_{a}u_{b}$, such that it projects into three-dimensional hypersurfaces-the observers' 3-D, instantaneous \textit{rest-space}-orthogonal to $u^{a}$ ($h_{ab}u^{b}=0$, $h^{a}{}_{a}=3$, $h_{a}{}^{c}h_{bc}=h_{ab}$). It is thus possible on using the $u^{a}$-field and its tensor counterpart $h_{ab}$ to split in a unique way any space-time variable, operator or equation in its temporal and spatial components. For instance, a given 4-vector field (e.g. consider the electromagnetic 4-potential $P^{a}$) decomposes in the following way
\begin{equation}
    P^{a}=\mathcal{P}u^{a}+\mathcal{P}^{a}\,,
\end{equation}
where $\mathcal{P}\equiv -P^{a}u_{a}$ is its (time-like) component which is parallel to the 4-velocity, and $\mathcal{P}^{a}\equiv h^{a}{}_{b}P^{b}\equiv P^{\langle a\rangle}$ is its projection into the 3-D hypersurfaces orthogonal to $u^{a}$. Similarly, a symmetric second-rank tensor field $T_{ab}$ can be split up as\footnote{The decomposition is based on the expansion of the expression $T_{ab}=g_{ac}g_{bd}T^{cd}=(h_{ac}-u_{a}u_{c})(h_{bd}-u_{b}u_{d})T^{cd}$.}
\begin{equation}
    T_{ab}=tu_{a}u_{b}+\frac{1}{3}(T^{c}{}_{c}+t)h_{ab}+2u_{ (a}t_{b)}+t_{ab}\,,
\end{equation}
where $t\equiv T_{ab}u^{a}u^{b}$, $t_{a}\equiv -h_{a}{}^{b}T_{bc}u^{c}$ and $t_{ab}\equiv h_{\langle a}{}^{c}h_{b\rangle}{}^{d}T_{cd}$.\footnote{Round brackets denote symmetrisation while square brackets imply antisymmetrisation. Angular brackets are used to describe the symmetric and trace-free part of an orthogonally projected second-rank tensor (e.g. $T_{\langle ab\rangle}=T_{(ab)}-(1/3)T^{c}{}_{c}h_{ab}$).} An example of such a second-rank tensor field is the energy-momentum tensor of a viscous fluid (refer to subsection~\ref{subsec:Fluid-des}).

Furthermore, the temporal and spatial derivatives of a general tensor field $T_{ab...}{}^{cd...}$ can be defined as
\begin{equation}
    \dot{T}_{ab...}{}^{cd...}\equiv u^{e}\nabla_{e}T_{ab...}{}^{cd...}
\end{equation}
and
\begin{equation}
    {\rm D}_{e}T_{ab...}{}^{cd...}\equiv h_{e}{}^{s}h_{a}{}^{f}h_{b}{}^{p}h_{q}{}^{c}h_{r}{}^{d}...\nabla_{s}T_{fp...}{}^{qr...}
\end{equation}
respectively, where $\nabla_{a}$ is the covariant differentiation operator of the 4-D space-time. Finally, let us define the totally antisymmetric 4-D Levi-Civita pseudotensor $\eta_{abcd}$ via the relations: $\eta_{abcd}\eta^{efpq}\equiv-4!\delta_{[a}{}^{e}\delta_{b}{}^{f}\delta_{c}{}^{p}\delta_{d]}{}^{q}$ and $\eta^{0123}\equiv[-\text{det}g_{ab}]^{-1/2}$. Now the 3-D Levi-Civita pseudotensor $\epsilon_{abc}$ is defined via the contraction of its 4-D counterpart along the time direction, $\epsilon_{abc}\equiv\eta_{abcd}u^{d}$. It follows that
\begin{equation}
    \epsilon_{abc}u^{a}=0 \quad \text{and} \quad \epsilon_{abc}\epsilon^{def}=3!h_{[a}{}^{d}h_{b}{}^{e}h_{c]}{}^{f}\,.
\end{equation}

\subsection{Kinematic quantities}

The motion of an observer with 4-velocity $u^{a}$ is characterised by a set of irreducible kinematic quantities which emerge from the decomposition of its velocity gradient into its symmetric trace-free part\footnote{Note that $\sigma_{ab}=D_{\langle b}u_{a\rangle}={\rm D}_{( b}u_{a)}-(1/3){\rm D}^{c}u_{c}h_{ab}$.}, its trace and its antisymmetric part,
\begin{equation}
    \nabla_{b}u_{a}=\sigma_{ab}+\omega_{ab}+\frac{1}{3}\Theta h_{ab}-\dot{u}_{a}u_{b}\,,
\end{equation}
where the sum of $\sigma_{ab}={\rm D}_{\langle b}u_{a\rangle}$, $\omega_{ab}={\rm D}_{[b}u_{a]}$ and $\Theta={\rm D}^{a}u_{a}$, namely of the shear and the vorticity tensors and the volume expansion/contraction scalar respectively, represents the spatial component of the 4-velocity gradient (${\rm D}_{b}u_{a}=\sigma_{ab}+\omega_{ab}+(1/3)\Theta h_{ab})$ which describes the relative motion of neighbouring observers. On the other hand, $\dot{u}_{a}u_{b}$ represents its temporal counterpart, where $\dot{u}^{a}=u^{b}\nabla_{b}u^{a}$ is the 4-acceleration vector. The presence of the latter is directly related to the existence of non-gravitational forces and therefore vanishes when the fluid moves along geodesic worldlines. By construction we have $\sigma_{ab}u^{b}=0=\omega_{ab}u^{b}=\dot{u}_{a}u^{a}$.

On using the 3-D Levi-Civita pseudotensor we can define the vorticity vector as $\omega^{a}=(1/2)\epsilon^{abc}\omega_{bc}$. In particular, the vorticity describes changes regarding the orientation of a given fluid element while the shear determines how the fluid's shape changes leaving its volume unaffected. Finally, the volume scalar refers to the average separation between neighbouring observers.

\subsection{Matter and electromagnetic fields}\label{Matter-EM-fields}

The dynamics of the matter fields is described by the well-known continuity and Euler's equations. Within the framework of general relativity these equations are derived from the zero divergence of the energy-momentum tensor, a consequence of the combined Einstein field equations and the Bianchi identities. As for the electromagnetic field dynamics, it is encoded by the familiar Maxwell equations. We present firstly the relativistic (covariant) versions of the equations in question. Secondly, we point out the unique coupling of the electromagnetic fields with spacetime curvature via the Ricci identities.

\subsubsection{Fluid description}\label{subsec:Fluid-des}

Both matter and electromagnetic fields accommodate a fluid description which is summarised in their energy-momentum tensor. The form of the latter depends on the physical properties of the fields as well as on the observer's coordinate frame. In the case of a viscous matter fluid the energy-momentum tensor reads
\begin{equation}
T_{ab}^{(\text{m})}=\rho u_{a}u_{b}+Ph_{ab}+2u_{(a}q_{b)}+\pi_{ab}\,,
\end{equation}
where $\rho=T_{ab}u^{a}u^{b}$ is the relativistic energy density (the rest mass density plus the total internal energy due to heat, chemical energy, etc.), $P=(h^{ab}/3)T_{ab}$ is the relativistic isotropic pressure, $q_{a}=-h_{a}{}^{b}T_{bc}u^{c}$ the energy flux relative to $u^{a}$ or the relativistic momentum density (due to diffusion or heat conduction), and $\pi_{ab}=h_{\langle a}{}^{c}h_{b\rangle}{}^{d}T_{cd}$ the relativistic anisotropic (trace-free) stress tensor (due to viscosity or free-streaming), all measured in the fundamental frame. Let us note that a perfect fluid model requires that $q_{a}=0=\pi_{ab}$.

Similarly, in the case of an electromagnetic fluid we have
\begin{equation}
    T_{ab}^{(\text{em})}=\frac{1}{2}(E^{2}+B^{2})u_{a}u_{b}+\frac{1}{6}(E^{2}+B^{2})h_{ab}+2\mathcal{Q}_{(a}u_{b)}+\Pi^{\text{(em)}}_{ab},
\end{equation}
where $E_{a}=F_{ab}u^{b}$ and $B_{a}=(1/2)\epsilon_{abc}F^{bc}$ represent the electric and the magnetic Maxwell field components respectively of the Faraday tensor,
\begin{equation}
    F_{ab}=2u_{[a}E_{b]}+\epsilon_{abc}B^{c}\,,
\end{equation}
as measured by a fundamental observer; $E^{2}=E^{a}E_{a}$ and $B^{2}=B^{a}B_{a}$ the square magnitudes of the individual fields, $\rho^{(\text{em})}=\frac{1}{2}(E^{2}+B^{2})$ is the energy density, $P^{(\text{em})}=\frac{1}{6}(E^{2}+B^{2})$ the isotropic pressure, $\mathcal{Q}_{a}=\epsilon_{abc}E^{b}B^{c}$ the Poynting vector or the electromagnetic energy flux and $\Pi_{ab}=-E_{\langle a}E_{b\rangle}-B_{\langle a}B_{b\rangle}$ the anisotropic pressure.\footnote{From the expression for $\Pi^{\text{(em)}}_{ab}$ it becomes evident that an electromagnetic fluid is necessarily viscous.}

Now the continuity equation as well as the equations of motion for a charged, conducting fluid are derived from the zero divergence condition (as implied by the combined Einstein's field equations and Bianchi identities) of the total energy-momentum tensor
\begin{equation}
\nabla^{b}T_{ab}=\nabla^{b}(T_{ab}^{(\text{em})}+T_{ab}^{(\text{m})})=0\,,
\label{eqn:total-cons-law}
\end{equation}
where $T_{ab}=T_{ab}^{(\text{em})}+T_{ab}^{(\text{m})}$ and\footnote{Equation~\eqref{div-of-EM-T} is derived with the aid of Maxwell's equations-see~\eqref{eqn:Maxwell1-2} in the following subsection.}
\begin{equation}
    \nabla^{b}T_{ab}^{(\text{em})}=-F_{ab}J^{b}
    \label{div-of-EM-T}
\end{equation}
with $J_{a}=\mu u_{a}+\mathcal{J}_{a}$ representing the electric 4-current, $\mu=-J^{a}u_{a}$ the electric charge and $\mathcal{J}_{a}=h_{a}{}^{b}J_{b}$ the orthogonally projected electric current. In particular, the timelike component of~\eqref{eqn:total-cons-law} (projection along $u^{a}$) leads to the continuity equation (or the energy conservation law)
\begin{equation}
    \dot{\rho}=-\Theta(\rho+P)-{\rm D}^{a}q_{a}-2\dot{u}^{a}q_{a}-\sigma^{ab}\pi_{ab}+E^{a}\mathcal{J}_{a}\,,
    \label{continuity-eq}
\end{equation}
which determines the rate of change of relativistic energy along the worldlines. It is worth noting that the above relativistic equation includes a term due to viscosity (the fourth one on its right-hand side), in remarkable contrast to its ordinary counterpart which is the same for any fluid model, whether viscous or not.

On the other hand, the spacelike component of~\eqref{eqn:total-cons-law} (projection orthogonal to $u^{a}$) leads to the equations of motion or Euler's equations (an expression of the momentum conservation law)
\begin{equation}
    (\rho+P)\dot{u}_{a}=-{\rm D}_{a}P-\dot{q}_{\langle a\rangle}-\frac{4}{3}\Theta q_{a}-(\sigma_{ab}+\omega_{ab})q^{b}-{\rm D}^{b}\pi_{ab}-\pi_{ab}\dot{u}^{b}+\mu E_{a}+\epsilon_{abc}\mathcal{J}^{b}B^{c}\,,
    \label{eqn:Euler's-full}
\end{equation}
which determines the acceleration caused by various pressure contributions. The sum $\rho+P$ describes the relativistic total inertial mass of the medium. The last two (electromagnetic) terms on the right-hand side of the above equation represent the familiar form of the Lorentz force.

\subsubsection{Maxwell equations}

The Maxwell equations are

\begin{equation}
    \nabla^{b}F_{ab}=J_{a} \quad \text{and} \quad \nabla_{[c}F_{ab]}=0\,.
    \label{eqn:Maxwell1-2}
\end{equation}
On using the definitions of the electric and magnetic field components presented in the previous subsection, the 1+3 split of Maxwell equations leads to a set of two propagation equations, these are
\begin{equation}
    \dot{E}_{\langle a\rangle}=-\frac{2}{3}\Theta E_{a}+(\sigma_{ab}+\epsilon_{abc}\omega^{c})E^{b}+\epsilon_{abc}\dot{u}^{b}B^{c}+\text{curl}B_{a}-\mathcal{J}_{a}\,,
    \label{eqn:el-field-prop}
\end{equation}
\begin{equation}
    \dot{B}_{\langle a\rangle}=-\frac{2}{3}\Theta B_{a}+(\sigma_{ab}+\epsilon_{abc}\omega^{c})B^{b}-\epsilon_{abc}\dot{u}^{b}E^{c}-\text{curl}E_{a}
    \label{eqn:magn-field-prop}
\end{equation}
and the following divergence conditions
\begin{equation}
    {\rm D}^{a}E_{a}+2\omega^{a}B_{a}=\mu\,,
    \label{el-div}
\end{equation}
\begin{equation}
    {\rm D}^{a}B_{a}-2\omega^{a}E_{a}=0\,.
    \label{magn-div}
\end{equation}
Equations~\eqref{eqn:el-field-prop},~\eqref{eqn:magn-field-prop},~\eqref{el-div} and~\eqref{magn-div} constitute 1+3 covariant versions of Amp\`ere's, Faraday's, Coulomb's and Gauss's law respectively.
For a set of Minkowski observers ($\dot{u}^{a}=\omega^{a}=\sigma_{ab}=\Theta=0$) the above equations reduce to the well-known form of Maxwell's equations.

Maxwell equations (see the first-the left one-set of equations in~\eqref{eqn:Maxwell1-2}) together with the antisymmetry of the Faraday tensor imply the zero divergence of the current 4-vector, $\nabla^{a}J_{a}=\nabla^{a}(\mu u_{a}+\mathcal{J}_{a})=0$, which leads to the electric charge conservation law
\begin{equation}
    \dot{\mu}=-\Theta\mu-{\rm D}^{a}\mathcal{J}_{a}-\dot{u}^{a}\mathcal{J}_{a}.
    \label{charge-conservation}
\end{equation}
In the absence of spatial currents, the temporal evolution of the charge density is determined by the volume scalar of the fluid.

\subsubsection{Matter-Electromagnetic fields and spacetime curvature}

Although a field theory describing both gravity and electromagnetism in a unified context is elusive, one can still study the interaction (or generally the coupling) between the spacetime curvature and the electromagnetic fields by incorporating the electromagnetic energy-momentum tensor in Einstein's field equations for gravity,
\begin{equation}
    R_{ab}-\frac{1}{2}Rg_{ab}=\kappa T_{ab}\,.
    \label{Einstein}
\end{equation}
In the above $R_{ab}$ is the (symmetric) Ricci tensor-encoding the local gravitational field, and $R=R^{a}_{a}$ is the Ricci scalar, which measures the mean local curvature. As we have seen in the previous subsections, the dynamical description of a fluid is achieved via the zero divergence of eq.~\eqref{Einstein}. 

Beyond this standard description of the various energy sources, the electromagnetic fields directly couple, due to their vector nature, with the spacetime curvature via the Ricci identities\footnote{In the context of our relativistic framework, we adopt a Riemannian spacetime model-with zero torsion}~\cite{T1, MavT, T3}
\begin{equation}
    2\nabla_{[a}\nabla_{b]}B_{c}=R_{abcd}B^{d}\,.
    \label{4-D-Ricci-id}
\end{equation}
The latter relation is written for the magnetic vector field and evidently a similar one holds for the electric component of the Maxwell field. The presence of the Riemann tensor $R_{abcd}$, which encodes the total gravitational field, on the right-hand side of the Ricci identities implies that the parallel transport of the vector $B_{a}$ from a given spacetime point to another depends on the geometric path followed. Note that this special status of the electromagnetic fields, owing to their vector nature, distinguishes them from all the other known energy sources, such as the ordinary matter.

On projecting equation~\eqref{4-D-Ricci-id} into the observer's 3-D, instantaneous rest-space, where measurements are made, we arrive at
\begin{equation}
    2{\rm D}_{[a}{\rm D}_{b]}B_{c}=-2\omega_{ab}\dot{B}_{\langle c\rangle}+\mathcal{R}_{dcba}B^{d}\,,
    \label{3-D-Ricci}
\end{equation}
where $\mathcal{R}_{dcba}$ is the associated 3-D Riemann tensor. In case that the fluid flow is irrotational (i.e. $\omega_{ab}=0$) the observers' 3-D tangent rest-planes form (integrable) hypersurfaces of simultaneity, orthogonal to their worldlines.

\section{Introducing a 1+2 split of the spatial components}\label{sec:1+2-split}

In some cases, a further 1+2 decomposition of the 3-dimensional space (leading to an overall 1+1+2 spacetime splitting--see~\cite{CB},~\cite{CMBD} and~\cite{GT} for some introductory information) in one specific spatial direction and a 2-dimensional surface orthogonal to it, may reveal additional useful information about the problem in hand. This is more likely to happen when the geometry, or the physics select a preferred spatial direction. For instance, one could consider the radial component of a spherically symmetric spacetime, or the rotation axis of a magnetised star, which may also happen to be parallel to the direction of the magnetic forcelines. However, a split of the spatial components may reveal valuable information about the problem in hand even there are not any apparent, favorable geometric or physical conditions (e.g. see the decomposition of Maxwell equations in the present piece of work.).

\subsection{Background}\label{ssec:1+2-backgr}

In what follows we show how 3-D mathematical objects (vectors, tensors, equations etc.) decompose into a component parallel to a spatial direction and two components lying on a 2-D surface perpendicular to the aforementioned direction~\cite{CB}. Let us introduce a space-like unit vector $n^{a}$ orthogonal to $u^{a}$ ($n^{a}n_{a}=1$, $n^{a}u_{a}=0$), which defines a specific spatial direction. Subsequently, we can define the symmetric tensor $\tilde{h}_{ab}\equiv h_{ab}-n_{a}n_{b}$ which projects vectors onto 2-D surfaces orthogonal to $n^{a}$ ($\tilde{h}_{ab}n^{b}=0$, $\tilde{h}^{a}{}_{a}=2$, $\tilde{h}_{a}{}^{c}\tilde{h}_{bc}=\tilde{h}_{ab}$). In analogy with the 1+3 formalism, 3-vectors and the corresponding second-rank, symmetric and trace-free tensors are split in their irreducible components according to the relations
\begin{equation}
    v^{a}=Vn^{a}+V^{a},
\end{equation}
where $V\equiv v^{a}n_{a}$ and $V^{a}\equiv \tilde{h}^{a}{}_{b}v^{b}$ while
\begin{equation}
    v_{ab}=V(n_{a}n_{b}-\frac{1}{2}\tilde{h}_{ab})+2V_{(a}n_{b)}+V_{ab},
\end{equation}
where $V\equiv v_{ab}n^{a}n^{b}=-\tilde{h}^{ab}v_{ab}$, $V_{a}\equiv \tilde{h}_{a}{}^{b}n^{c}v_{bc}$ and $V_{ab}\equiv (\tilde{h}_{(a}{}^{c}\tilde{h}_{b)}{}^{d}-(1/2)\tilde{h}_{ab}\tilde{h}^{cd})v_{cd}$.
For instance, let us consider the 1+1+2 decomposition of the energy-momentum tensor $T_{ab}=g_{ac}g_{bd}T^{cd}=(\tilde{h}_{ac}-u_{a}u_{c}+n_{a}n_{c})(\tilde{h}_{bd}-u_{b}u_{d}+n_{b}n_{d})$, which leads to
\begin{equation}
    T_{ab}=\rho u_{a}u_{b}+\tilde{\rho}n_{a}n_{b}+\tilde{P}\tilde{h}_{ab}+2u_{(a}q_{b)}+2n_{(a}\tilde{q}_{b)}+\Pi_{ab},
\end{equation}
where $\tilde{\rho}\equiv T_{ab}n^{a}n^{b}=P+\Pi$ and $\tilde{P}\equiv (\tilde{h}^{ab}/2)T_{ab}=P-\Pi/2$ (therefore $\Pi=(2/3)(\tilde{\rho}-\tilde{P}$)) are the analogues of relativistic energy density and pressure defined in reference to spacelike curves with tangent vector $n^{a}$. Regarding $\tilde{q}_{a}\equiv \tilde{h}_{a}{}^{b}n^{c}T_{bc}= \Pi_{a}$ and $\Pi_{ab}\equiv (\tilde{h}_{(a}{}^{c}\tilde{h}_{b)}{}^{d}-(1/2)\tilde{h}_{ab}\tilde{h}^{cd})T_{cd}$, they represent the (2-D) surface (normal to $n^{a}$) counterparts of the energy flux vector and the viscosity tensor respectively (refer to equation~\eqref{pi-decomp} for the decomposition of the anisotropic stress tensor).
We gather here for reference all of the decomposition relations of vectors and tensors, which we use throughout this article\footnote{Note that $\dot{n}_{a}n^{a}=0$ in eq.~\eqref{dot-n-dec} and therefore $\alpha_{a}n^{a}=0$.}
\begin{equation}
    \dot{u}^{a}=\mathcal{A}n^{a}+\mathcal{A}^{a}
\end{equation}
\begin{equation}
\dot{n}^{a}=\mathcal{A}u^{a}+\alpha^{a}
\label{dot-n-dec}
\end{equation}
\begin{equation}
    \omega^{a}=\Omega n^{a}+\Omega^{a}
\end{equation}
\begin{equation}
    q^{a}=Qn^{a}+Q^{a}
\end{equation}
\begin{equation}
    E^{a}=\epsilon n^{a}+\epsilon^{a}
\end{equation}
\begin{equation}
    B^{a}=\mathcal{B}n^{a}+\mathcal{B}^{a}
\end{equation}
\begin{equation}
    \mathcal{J}^{a}=jn^{a}+j^{a}
\end{equation}
\begin{equation}
    \sigma_{ab}=\Sigma(n_{a}n_{b}-\frac{1}{2}\tilde{h}_{ab})+2\Sigma_{(a}n_{b)}+\Sigma_{ab}
    \label{sigma-1+2}
\end{equation}
\begin{equation}
    \pi_{ab}=\Pi(n_{a}n_{b}-\frac{1}{2}\tilde{h}_{ab})+2\Pi_{(a}n_{b)}+\Pi_{ab}
    \label{pi-decomp}
\end{equation}
\begin{equation}
    E_{ab}=\mathcal{E}(n_{a}n_{b}-\frac{1}{2}\tilde{h}_{ab})+2\mathcal{E}_{(a}n_{b)}+\mathcal{E}_{ab}\,.
\end{equation}
In the last equation, $E_{ab}$ is the electric component of the Weyl (long-range) curvature tensor. There is also the magnetic tensor component $H_{ab}$. Weyl curvature is associated with tidal forces and gravitational waves (e.g. refer to~\cite{TCM}). The aforementioned decomposition relation will be used only once when discussing the gravitational collapse of a magnetised fluid in section~\ref{sec:Grav-collapse}. Finally, for some details concerning the meaning of the shear's individual components see the appendix section~\ref{Shear-1+2-comp}. 

Regarding the derivatives of a general tensor field $T_{ab...}{}^{cd...}$, the one along $n^{a}$ and the other projected on the 2-surface normal to $n^{a}$, these are defined respectively as
\begin{equation}
    T'_{ab...}{}^{cd...}\equiv n^{e}{\rm D}_{e}T_{ab...}{}^{cd...}
\end{equation}
and
\begin{equation}
    \tilde{{\rm D}}_{e}T_{ab...}{}^{cd...}\equiv \tilde{h}_{e}{}^{s}\tilde{h}_{a}{}^{f}\tilde{h}_{b}{}^{p}\tilde{h}_{q}{}^{c}\tilde{h}_{r}{}^{d}...{\rm D}_{s}T_{fp...}{}^{qr...}\,.
\end{equation}
Finally, the 2-D Levi-Civita pseudotensor can be defined via the contraction of its 3-D counterpart along the spatial direction of $n^{a}$, $\epsilon_{ab}\equiv\epsilon_{abc}n^{c}$. It follows that
\begin{equation}
    \epsilon_{ab}n^{b}=0 \quad \text{and} \quad \epsilon_{ab}\epsilon^{cd}=2\tilde{h}_{[a}{}^{c}\tilde{h}_{b]}{}^{d}
\end{equation}
as well as that $\epsilon_{abc}=n_{a}\epsilon_{bc}-n_{b}\epsilon_{ac}+n_{c}\epsilon_{ab}$.

\subsection{Kinematic quantities}

In analogy with its 3-D counterpart the motion on the 2-D surface orthogonal to $n^{a}$ is characterised by a set of kinematic quantities which come from the decomposition of the gradient of $n^{a}$. In other words, we have
\begin{equation}
    {\rm D}_{b}n_{a}=\tilde{\sigma}_{ab}+\tilde{\omega}_{ab}+\frac{1}{2}\tilde{\Theta}\tilde{h}_{ab}+n_{a}n'_{b},
\end{equation}
where $\tilde{\sigma}_{ab}\equiv {\rm D}_{\langle b}n_{a\rangle}$, $\tilde{\omega}_{ab}\equiv {\rm D}_{[b}n_{a]}$ and $\tilde{\Theta}\equiv {\rm D}^{a}n_{a}$ are respectively the shear and the vorticity tensors, the surface expansion-contraction scalar and $n'_{a}\equiv n^{b}{\rm D}_{b}n_{a}$ the spatial derivative of $n^{a}$ along its own direction. The sum $\tilde{{\rm D}}_{b}n_{a}=\tilde{\sigma}_{ab}+\tilde{\omega}_{ab}+\frac{1}{2}\tilde{\Theta}\tilde{h}_{ab}$ describes the relative motion of neighbouring spacelike curves orthogonal to the surface in question.

We encourage the reader to compare the 2-D version of the shear $\tilde{\sigma}_{ab}\equiv {\rm D}_{\langle b}n_{a\rangle}$ with those of the individual 1+2 components of its 3-D version $\sigma_{ab}\equiv {\rm D}_{\langle b}u_{a\rangle}$ found in Appendix~\ref{Shear-1+2-comp}. Concerning the 2-D vorticity tensor, it has only one independent component (i.e. it consists of a vector along the one of the two independent directions defining the 2-D surface), so that it can be written as $\tilde{\omega}_{ab}=\tilde{\omega}\epsilon_{ab}$, where $\tilde{\omega}^{2}=(1/2)\tilde{\omega}^{ab}\tilde{\omega}_{ab}$. Finally, the condition $n'^{a}=0$ implies that the $n^{a}$ field is tangent to a congruence of spacelike geodesics.

\subsection{1+1+2 System of equations for a magnetised fluid}

Within the framework of ordinary electrodynamics of continuous media~\cite{LLP} (where Newtonian instead of relativistic gravity is adopted), the description of a conducting fluid in a magnetic field requires, on the one hand, the fluid dynamics equations, namely the continuity equation, Euler's equation of motion and an equation of state\footnote{In general, the equation of state relates the pressure, density and temperature of the fluid, $P=P(\rho, T)$. The dependence on the temperature requires the equation of heat transfer for the system to be completed. However, for our purposes a barotropic equation of state, $P=w\rho$ with $w=\text{const.}$ will be sufficient.}; on the other hand, Maxwell's electrodynamic field equations.

Regarding our relativistic approach, the whole Einstein-Maxwell system of equations (which includes the long range or Weyl gravitational fields as well) is generally needed to fully describe the motion of a magnetised fluid. Nevertheless, as our interest focuses particularly on the behavior or the evolution of the magnetic field and its implications on the motion of the fluid, we will eliminate our attention to the Euler-Maxwell system of equations. Besides, it turns out that the 1+2 decomposed Euler-Maxwell system of equations at the ideal MHD limit does not involve directly the effects of the long range gravitational field.

In the following subsections we firstly consider the ideal MHD limit of the system in question and subsequently split up its equations in their 1+2 spatial components. We conduct our calculations by defining the arbitrary spacelike vector $n^{a}$, which we use for making the 1+2 decomposition, to be parallel to the magnetic field lines. The 1+2 split of the full equations as well as argumentation showing the equivalence of the system under the alternative assumption $B^{a}\perp n^{a}$, are included for the interested reader in Appendix~\ref{app:1+2-full} and~\ref{sec:Equiv-system} respectively.

\subsubsection{The magnetohydrodynamics (MHD) approximation}

Aiming to describe the motion of a magnetised fluid, we need to isolate the magnetic component of the Maxwell field. This can be achieved theoretically by adopting a highly conducting fluid model. According to Ohm's law applied in the fluid's rest frame,
\begin{equation}
    \mathcal{J}_{a}=\varsigma E_{a}\,,
    \label{Ohm}
\end{equation}
non-zero spatial currents arise for $E_{a}\rightarrow 0$ at the MHD limit (i.e. $\varsigma\rightarrow\infty$, where $\varsigma$ is the conductivity of the medium). For such a perfect conductor the magnetic field lines behave as being \textit{frozen} in the fluid.

\subsubsection{Magnetic field equations and solution}\label{ssec:magn-field-eqs}

Making use of the MHD approximation, Maxwell's equations~\eqref{eqn:el-field-prop}-\eqref{magn-div} reduce to one propagation equation
\begin{equation}
    \dot{B}_{\langle a\rangle}=(-\frac{2}{3}\Theta h_{ab}+\sigma_{ab}+\epsilon_{abc}\omega^{c})B^{b}\,,
    \label{MHD-B-prop}
\end{equation}
known as the \textit{magnetic induction equation}, which shows that the temporal evolution of the magnetic field is a direct result of the relative motion of neighbouring fluid particles; and three constraints
\begin{equation}
    \mathcal{J}_{a}=\epsilon_{abc}\dot{u}^{b}B^{c}+\text{curl}B_{a}\,,
    \label{MHD-current}
\end{equation}
\begin{equation}
    \omega^{a}B_{a}=\mu \quad \text{and} \quad {\rm D}^{a}B_{a}=0\,,
    \label{MHD-constraints}
\end{equation}
where according to~\eqref{MHD-current} the magnetic field lines remain frozen--in with the matter, in the form of currents. Subsequently, projecting Faraday's law, eq.~\eqref{MHD-B-prop}, along and orthogonal to an arbitrary spacelike vector $n^{a}$, defined along the direction of the field lines (i.e. $B^{a}=\mathcal{B}n^{a}$), we arrive at
\begin{equation}
    \dot{\mathcal{B}}=-\Theta \mathcal{B} \hspace{15mm} \text{and} \hspace{15mm} \alpha_{a}=-2\epsilon_{ac}\Omega^{c}=u'_{a}\,,
    \label{eqn:dot-beta+alpha-constr}
\end{equation}
where we have taken into account the decomposition relations in section~\ref{ssec:1+2-backgr} as well as that $\Sigma=-\Theta/3$ and $\Sigma_{a}=-\epsilon_{ab}\Omega^{b}$ (see the Appendix~\ref{Shear-1+2-comp}). We observe that eq.~(\ref{eqn:dot-beta+alpha-constr}a) is a covariant, linear, partial differential equation of first order. It appears that our decomposition has brought the Faraday's law into a solvable form. The latter tells us that the rate of change of the magnetic field along the worldlines is proportional to the expansion or contraction of a given volume containing the worldlines. In the following we provide a general method of solving differential equations of the form in question. We will proceed to the solution after writing down the decomposed constraint relations for the magnetic field. In particular, eq.~\eqref{MHD-current} splits into
\begin{equation}
-\mathcal{B}^{2}\mathcal{A}_{a}-2\mathcal{B}\tilde{\rm{D}}_{a}\mathcal{B}+\mathcal{B}^{2}n'_{a}=\mathcal{B}\epsilon_{ac}j^{c} \hspace{15mm} \text{and} \hspace{15mm} \tilde{\omega}\mathcal{B}=-\frac{j}{2}\,.
\label{eqn:bj}
\end{equation}
As for the scalar equations~\eqref{MHD-constraints}, they are written as
\begin{equation}
\Omega \mathcal{B}=\frac{\mu}{2}\hspace{15mm} \text{and} \hspace{15mm} \mathcal{B}'+\tilde{\Theta}\mathcal{B}=0\,.
\label{b'-Omega-b}
\end{equation}
Both the charge density $\mu$ and the current along the magnetic forcelines $j$ are determined by the magnetic field $\mathcal{B}$ and the value of the vorticity vector along and orthogonal to $B^{a}$ respectively. Moreover, note the remarkable similarity between equations~(\ref{eqn:dot-beta+alpha-constr}a) and~(\ref{b'-Omega-b}b), namely the decomposed forms of Faraday's and Gauss' law respectively.

In what follows, we proceed to the solution of~(\ref{eqn:dot-beta+alpha-constr}a) which provides the paradigm for the solution of~(\ref{b'-Omega-b}b). First of all, as $\mathcal{B}$ is a scalar quantity, its covariant differentiation is equivalent to its ordinary differentiation, so that
\begin{equation}
    \dot{\mathcal{B}}=u^{a}\nabla_{a}\mathcal{B}=u^{a}\partial_{a}\mathcal{B}=(u^{0}\partial_{0}+u^{1}\partial_{1}+u^{2}\partial_{2}+u^{3}\partial_{3})\mathcal{B}=-\Theta \mathcal{B}\,.
    \label{eqn:bdot}
\end{equation}
Now by defining new space-time variables $\tilde{x}^{a}$ such that\footnote{Note that here the repeated index $i$ does not imply summation of components.}
\begin{equation}
    \tilde{x}^{i}=\int\frac{dx^{i}}{u^{i}}\,,
\end{equation}
expression~\eqref{eqn:bdot} becomes
\begin{equation}
    (\tilde{\partial}_{0}+\tilde{\partial}_{1}+\tilde{\partial}_{2}+\tilde{\partial}_{3})\mathcal{B}=-\Theta \mathcal{B},
    \label{tilde-b-dot}
\end{equation}
where $\tilde{\partial}_{i}$ are the new derivative operators with respect to the variables $\tilde{x}^{i}$. Let us try to solve the latter equation by assuming variables separation : $\mathcal{B}=\mathcal{T}(\tilde{x}^{0})U(\tilde{x}^{1})V(\tilde{x}^{2})W(\tilde{x}^{3})$, where $\tilde{x}^{0}$ is the new temporal variable and $\tilde{x}^{1},\tilde{x}^{2},\tilde{x}^{3}$ are the new spatial variables. Relation~\eqref{tilde-b-dot} takes thus the form
\begin{equation}
    \frac{\tilde{\partial}_{0}\mathcal{T}}{\mathcal{T}}+\frac{\tilde{\partial}_{1}U}{U}+\frac{\tilde{\partial}_{2}V}{V}+\frac{\tilde{\partial}_{3}W}{W}=-\Theta(\tilde{x}^{0},\tilde{x}^{1},\tilde{x}^{2},\tilde{x}^{3}),
    \label{sep-variables2}
\end{equation}
We observe that each of the fractions in the above equation depends only on one of the variables $\tilde{x}^{i}$. Subsequently, equation~\eqref{sep-variables2} holds if and only if $\Theta(\tilde{x}^{0},\tilde{x}^{1},\tilde{x}^{2},\tilde{x}^{3})=\Theta_{0}(\tilde{x}^{0})+\Theta(\tilde{x}^{1})+\Theta_{2}(\tilde{x}^{2})+\Theta_{3}(\tilde{x}^{3})$. Therefore, the original partial differential equation reduces to four ordinary differential equations of the form $(\tilde{\partial}_{1}U/U)=-\Theta_{1}(\tilde{x}^{1})$, which are integrated directly to give $U=c_{1}e^{-\int{\Theta_{1}}d\tilde{x}^{1}}$. Hence, it is overall clear to see that the solution for $\mathcal{B}$ can be written as
\begin{equation}
\mathcal{B}=\mathcal{C}e^{-\int{\Theta_{0}}d\tilde{x}^{0}-\int{\Theta_{1}}d\tilde{x}^{1}-\int{\Theta_{2}}d\tilde{x}^{2}-\int{\Theta_{3}}d\tilde{x}^{3}}=\mathcal{C}e^{-\int{\frac{\Theta_{0}}{u^{0}}}dx^{0}-\int{\frac{\Theta_{1}}{u^{1}}}dx^{1}-\int{\frac{\Theta_{2}}{u^{2}}}dx^{2}-\int{\frac{\Theta_{3}}{u^{3}}}dx^{3}}\,,
\label{eqn:bexp}
\end{equation}
where $\mathcal{C}$ is an arbitrary constant and we have found out that our variables separation assumption turns out to be true\footnote{Recall that the original equation~\eqref{eqn:dot-beta+alpha-constr} is a partial differential one. However, we have shown that it reduces four ordinary equations (see~\eqref{sep-variables2}). As a consequence, the general solution we have found, eq.~\eqref{eqn:bexp} is actually the only solution of the original equation.}. Equation~\eqref{eqn:bexp}, which is a solution\footnote{As far as we know, it is the first time that the solution in question appears in the literature.} of Faraday's law at the MHD limit, tells us that if $\Theta_{i}(\tilde{x}^{i})$ are continuous functions in a specific closed interval $[\alpha_{1}, \alpha_{2}]$ of their domain and they preserve constant sign (e.g. $\Theta_{i}(\tilde{x}^{i})\leq 0$--implying continuous gravitational contraction) for every value of their variable belonging in the interval, then $\int_{\alpha_{1}}^{\alpha_{2}}{ \Theta_{i}(\tilde{x}^{i})}d\tilde{x}^{i}<0$ and the magnetic field generally obeys an exponential type of increase with respect to the spacetime variables. In fact, the aforementioned exponential type behavior seems to be outward because on defining a scale factor $a(\tilde{x}^{0},\tilde{x}^{1},\tilde{x}^{2},\tilde{x}^{3})$, such that $\Theta=3\dot{a}/a$ (also $\Theta_{0}=3{\rm d}a_{0}/(a_{0}{\rm d}\tilde{x}^{0})$ and $\Theta_{i}=3{\rm d}a_{i}/(a_{i}{\rm d}\tilde{x}^{i}))$, equation~\eqref{eqn:bexp} reduces to
\begin{equation}
    \mathcal{B}\propto a^{-3}=\left(a_{0}(\tilde{x}^{0})a_{1}(\tilde{x}^{1})a_{2}(\tilde{x}^{2})a_{3}(\tilde{x}^{3})\right)^{-3}\,.
    \label{Bexp->B-a}
\end{equation}
Finally, we shall keep in mind the following remarks. Firstly, on deriving relations~\eqref{eqn:bexp},~\eqref{Bexp->B-a} we have not adopted a specific coordinate reference frame. Secondly, the evolution of $\mathcal{B}$ in each spacetime direction is independent of its evolution in the other directions with respect to the tilted variables only, where $\mathcal{B}=\mathcal{T}(\tilde{x}^{0})U(\tilde{x}^{1})V(\tilde{x}^{2})W(\tilde{x}^{3})$. The crucial equation~\eqref{eqn:bexp}, or~\eqref{Bexp->B-a}, provides us the keystone for studying magnetic fields in cosmological and astrophysical problems (refer to the following sections).

In order to specify the constant $\mathcal{C}$, we observe that the key fluid dynamic quantity related to the magnetic field, is the volume scalar $\Theta$. Therefore, we turn our attention to the relation which describes its evolution, the so-called Raychaudhuri equation (e.g. see~\cite{EMM}),
\begin{equation}
    \dot\Theta=-\frac{1}{3}\Theta^{2}-\frac{1}{2}(\rho+3P+\mathcal{B}^{2})-2(\sigma^{2}-\omega^{2})+D^{a}\dot{u}_{a}+\dot{u}^{a}\dot{u}_{a}\,.
    \label{Raych-eqn1}
\end{equation}
Considering an instant during which the fluid is found in its equilibrium state\footnote{Such an instant could have been either the initial instant-just before the collapse starts-or a transitional instant, during which the collapse stops and the fluid starts expanding.} (setting $\Theta=0=\sigma^{2}$ and $\dot u_{a}=0=\omega^{2}$), we have $\mathcal{B}=\mathcal{C}$, and~\eqref{Raych-eqn1} leads to (the star index refers to equilibrium values in the following)
\begin{equation}
    \mathcal{C}^{2}=-(2\dot\Theta_{*}+\rho_{*}+3P_{*}),
    \label{eqn:C1}
\end{equation}
which means that $\mathcal{C}$ is a real constant if
\begin{equation}
    \dot\Theta_{*}<-\frac{1}{2}(\rho_{*}+3P_{*})<0.
\end{equation}
In other words, the rate of change of the volume scalar in the equilibrium has to be negative and smaller than the gravitational mass of the system due to conventional matter ($\frac{1}{2}(\rho_{*}+3P_{*})>0$).

In the same way eq.~(\ref{b'-Omega-b}b) solves to give
\begin{equation}
    \mathcal{B}=\mathcal{F}e^{-\int{\frac{\tilde{\Theta}_{0}}{n^{0}}}dx^{0}-\int{\frac{\tilde{\Theta}_{1}}{n^{1}}}dx^{1}-\int{\frac{\tilde{\Theta}_{2}}{n^{2}}}dx^{2}-\int{\frac{\tilde{\Theta}_{3}}{n^{3}}}dx^{3}}\,,
    \label{b-exp3}
\end{equation}
where $\mathcal{F}$ is a constant. According to the latter relation, the magnetic field changes with the area scalar $\tilde{\Theta}$ (which describes the expansion/contraction of the 2-D surface orthogonal to the magnetic forcelines) in complete analogy with its dependence on the volume scalar $\Theta$. Note that the area scalar splits in components, $\tilde{\Theta}=\tilde{\Theta}_{0}(\tilde{x}^{0})+\tilde{\Theta}_{1}(\tilde{x}^{1})+\tilde{\Theta}_{2}(\tilde{x}^{2})+\tilde{\Theta}_{3}(\tilde{x}^{3})$, in full correspondence with its 3-D counterpart.

\subsubsection{Fluid dynamic equations}

At the ideal MHD limit ($q_{a}=0=\pi_{ab}$ and $E_{a}=0$), the equation of continuity~\eqref{continuity-eq} reduces to
\begin{equation}
    \dot{\rho}=-\Theta(\rho+P)\,.
    \label{continuity-eq2}
\end{equation}
It is worth noting that even if we had considered an imperfect (viscous) fluid model, the magnetic field would behave according to the same law--relation~\eqref{eqn:bexp} would still be true because equation~(\ref{eqn:dot-beta+alpha-constr}a) would have remained essentially the same. However, in that case, the constant $\mathcal{C}$ would have been given by a far more complicated expression while in general the comprehension as well as the application of the system to realistic problems (see the last two sections) would have been a far more difficult task.

Subsequently, assuming a barotropic equation of state of the form
\begin{equation}
    P=w\rho\,,
    \label{eq-of-state}
\end{equation}
where $0\leq w\leq 1$ is a constant parameter, the continuity equation finally becomes
\begin{equation}
    \dot{\rho}=-\Theta(1+w)\rho\,.
    \label{continuity-eq3}
\end{equation}
The latter shows that changes in the volume scalar determine the evolution of the matter density. In complete analogy with~(\ref{eqn:dot-beta+alpha-constr}a) and~(\ref{b'-Omega-b}b), equation~\eqref{continuity-eq3} solves to give
\begin{equation}
    \rho=\mathcal{D}e^{-\int{(1+w)\frac{\Theta_{0}}{u^{0}}}dx^{0}-\int{(1+w)\frac{\Theta_{1}}{u^{1}}}dx^{1}-\int{(1+w)\frac{\Theta_{2}}{u^{2}}}dx^{2}-\int{(1+w)\frac{\Theta_{3}}{u^{3}}}dx^{3}},
    \label{dens-evol2}
\end{equation}
where $\mathcal{D}$ is a constant. According to the above relation, in the case of dust (i.e. $w=0$), the density of matter evolves in the same way as the magnetic field does. On the other hand, the density of stiff matter (i.e. $w=1$) evolves in the same rate as the magnetic energy density $\mathcal{B}^{2}$ does.

Concerning Euler's equation, the application of the ideal MHD approximation leads to
\begin{equation}
    (\rho+P)\dot{u}_{a}=-{\rm D}_{a}P+\epsilon_{abc}\mathcal{J}^{b}B^{c}\,,
    \label{Euler-ideal-MHD}
\end{equation}
where the pressure gradients and the magnetic Lorentz force are the remaining causes of non-geodesic motion. Substituting the current from~\eqref{MHD-current} into the last term in the above relation and following the operations we arrive at
\begin{equation}
    \epsilon_{abc}\mathcal{J}^{b}B^{c}=-B^{2}\dot{u}_{a}+\dot{u}^{b}B_{b}B_{a}-\frac{1}{2}{\rm D}_{a}B^{2}+B^{b}{\rm D}_{b}B_{a}\,.
    \label{eqn:Lorentz-force}
\end{equation}
The last two terms in the right-hand side of the above relation are due to the magnetic pressure and the magnetic tension respectively. Therefore, eq.~\eqref{Euler-ideal-MHD} transforms into
\begin{equation}
    (\rho+P+B^{2})\dot{u}_{a}=-{\rm D}_{a}P+\dot{u}^{b}B_{b}B_{a}-\frac{1}{2}{\rm D}_{a}B^{2}+B^{b}{\rm D}_{b}B_{a}\,.
    \label{Euler-ideal-MHD-2}
\end{equation}
On projecting the above relation along and normal to $n^{a}$, it decomposes into
\begin{equation}
    (\rho+P)\mathcal{A}=-P'\hspace{15mm} \text{and} \hspace{15mm} (\rho+P+\mathcal{B}^{2})\mathcal{A}_{a}=-\tilde{{\rm D}}_{a}P-\mathcal{B}\tilde{{\rm D}}_{a}\mathcal{B}+\mathcal{B}^{2}n'_{a}\,.
\label{Euler-1+2}
\end{equation}
respectively. Not surprisingly, the motion along the magnetic field lines (eq.~(\ref{Euler-1+2}a)) is not determined by the effect of magnetic forces. As for the motion orthogonal to the field lines (eq.~(\ref{Euler-1+2}b)), it is determined not only by the associated pressure gradient but the magnetic pressure and tension as well.\footnote{Note that the equation of motion~(\ref{Euler-1+2}b) in the equilibrium state is written as
\begin{equation}
\mathcal{C}^{2}n'_{a*}=\tilde{{\rm D}}_{a}P_{*}.
\label{eqn:C2}
\end{equation}
Combining~\eqref{eqn:C1} and~\eqref{eqn:C2} one determines the value of $n'_{a}$ in the equilibrium,
\begin{equation}
    n'_{a*}=-\frac{\tilde{{\rm D}}_{a}P_{*}}{2\dot\Theta_{*}+\rho_{*}+3P_{*}}\,.
\end{equation}} Now taking into account the equation of state~\eqref{eq-of-state}, the individual components of Euler's equation transform into
\begin{equation}
    \mathcal{A}=\left(\ln\rho^{-\frac{1}{1+w}}\right)'\hspace{15mm} \text{and} \hspace{15mm} \mathcal{A}_{a}=-\frac{w\tilde{{\rm D}}_{a}\rho}{(1+w)\rho+\mathcal{B}^{2}}-\frac{\tilde{{\rm D}}_{a}\mathcal{B}^{2}}{2\left[(1+w)\rho+\mathcal{B}^{2}\right]}+c^{2}_{\mathcal{A}}n'_{a}\,,
    \label{Euler-1+2-2}
\end{equation}
where $c^{2}_{A}=\frac{\mathcal{B}^{2}}{\rho+P+\mathcal{B}^{2}}$ represents the square of the Alfv$\acute{\text{e}}$n velocity. In the next step, substituting the density evolution formula~\eqref{dens-evol2} into~(\ref{Euler-1+2-2}a) and following the operations, we finally arrive at
\begin{equation}
    \mathcal{A}=\frac{n^{1}\Theta_{1}}{u^{1}}+\frac{n^{2}\Theta_{2}}{u^{2}}+\frac{n^{3}\Theta_{3}}{u^{3}}\,,
    \label{A-cond}
\end{equation}
which shows in a direct manner that the motion along the magnetic forcelines is determined by the fluid's volume expansion or contraction. Regarding the motion orthogonal to the magnetic forcelines (see eq.~\ref{Euler-1+2}b)),--recalling the evolution of $\mathcal{B}$ and $\rho$--it appears that the magnetic force terms tend to dominate over the pressure or matter density gradient in the case of contraction ($\Theta<0$) whilst the opposite is expected to happen in the case of expansion ($\Theta>0$). This observation is based on a comparison of the exponential terms related to $\rho$ and $\mathcal{B}$. However, the exact behavior of the magnetic pressure term depends on the evolution of the $\Theta$ coefficients which come from the differentiation of $\mathcal{B}^{2}$. Besides, an exception to the aforementioned observation we have when considering a stiff matter model ($w=1$). In the last case both matter and magnetic energy densities evolve at the same rate.

\section{Cosmological magnetic fields in homogeneous models}\label{sec:Cosmol-magn-fields}

In this section we make use of equation~\eqref{eqn:bexp} with the aim of studying the evolution of large-scale magnetic fields. In the first place, we explain why the cosmic medium is expected to satisfy the ideal magnetohydrodynamic (MHD) requirements, which entail the subsequent application of~\eqref{eqn:bexp} in homogeneous and anisotropic cosmological spacetimes. In the second place, we focus on the Bianchi I model, the case of which provides a specific, indirect but clear verification of our general result within the literature. In particular, taking into account the magnetic energy contribution, we derive the evolution formulae of a Bianchi I model with perfect fluid content. Finally, we determine the epoch of equality between magnetic energy density and radiation/matter, considering in parallel the nucleosynthesis constraint in relation to the magnetic density evolution, within the model in question. Our estimation of the aforementioned equality epoch could fortunately be used as a reference point when studying the origin of cosmic magnetic fields in the pre-recombination era.

\subsection{The MHD approximation of the cosmic medium}\label{MHD-cosmic-medium}

Within the context of the standard cosmological model, large-scale gravitational as well as electromagnetic perturbations are causally produced via the inflationary mechanism. In particular, spacetime distortions initially appear in the form of quantum fluctuations during the so-called \textit{Planck epoch}. Subsequently, due to the exponential expansion of the \textit{inflation} era, these quantum fluctuations are forced to pass out of the Hubble horizon, where they freeze out in the form of classical perturbations. After inflation, during reheating and the following radiation era, the electrical conductivity of the initially poorly conducting cosmic medium increases rapidly\cite{EMM}. As a consequence, the electric fields gradually vanish and the currents freeze the magnetic fields in with the cosmic fluid. In other words, the post-inflationary universe can be causally described by the ideal magnetohydrodynamical model, within the Hubble scale. Besides, the adoption of the MHD approximation in the standard cosmological framework is in accordance with the fact that only large-scale magnetic (not electric) fields have been observed. In the following, our interest focuses on the evolution of large-scale magnetic fields lying within the Hubble horizon.

\subsection{Homogeneous anisotropic models hosting large--scale magnetic fields}

Let us consider the application of equation~\eqref{eqn:bexp}--recall that this relation requires that the MHD approximation is satisfied--in homogeneous and (expanding/contracting) anisotropic, cosmological spacetime. It simplifies to
\begin{equation}
    \mathcal{B}=\mathcal{C}e^{-\int{\frac{\Theta_{0}}{u^{0}}}dx^{0}}\,.
\end{equation}
We should note that the presence of the magnetic fields (defining a preferable spatial direction) presupposes or requires a certain anisotropy of their host cosmological environment. On using comoving (unchanged by the cosmic expansion) coordinates along the fundamental worldlines ($u^{0}=1$, $u^{i}=0$ and $x^{0}\rightarrow \tau$, where $\tau$ is the fundamental observer's proper time) and taking into account the definition of the Hubble parameter ($\Theta_{0}=3H=3\dot{a}/a$, where $a$ represents the average scale factor of the anisotropic spacetime), the above expression becomes
\begin{equation}
    \mathcal{B}=\mathcal{C}e^{-3\int\frac{da}{a}}=\mathcal{C}e^{-3\ln{a}}\rightarrow \mathcal{B}\propto a^{-3}~(x^{0}\equiv\tau)\,,
    \label{magn-field-BianchiI}
\end{equation}
so that the magnetic energy density satisfies
\begin{equation}
    \rho_{B}\propto \mathcal{B}^{2}\propto a^{-6}~(x^{0}\equiv\tau)\,.
    \label{magn-density-Bianchi}
\end{equation}
The validity of the above relation is restricted to homogeneous and anisotropic cosmological models which are able to accommodate pure, large-scale magnetic fields. It is known that of the so-called (homogeneous) \textit{Bianchi models}, there are some which potentially behave as natural hosts of large-scale magnetic fields. In particular, these are Bianchi I, II, III, $\text{VI}_{-1}$ and $\text{VII}_{0}$ in accordance with~\cite{HJ}. Note that equation~\eqref{magn-field-BianchiI} involves a significantly faster change of magnetic fields with time in comparison to their evolution in perturbed FRW models with flat spatial sections. Recall that in the latter case, the more familiar relation $\mathcal{B}\propto a^{-2}$ holds instead (e.g. see~\cite{MavT, T3}). The reader can refer to subsection~\ref{ssec:FRW-Bianchi} for a comparison regarding the relative evolution of magnetic fields and radiation/dust in perturbed Friedmann-Robertson-Walker and Bianchi I cosmological models.

\subsection{The Bianchi I case}\label{ssec:BianchiI}

Now we focus our attention specifically on the simplest anisotropically expanding cosmological model, namely the so-called Bianchi I, which has Euclidean spatial sections and is known to allow for the existence of  large-scale magnetic fields. Its metric in comoving coordinates reads
\begin{equation}
    ds^{2}=-dt^{2}+A^{2}(t)dx^{2}+B^{2}(t)dy^{2}+C^{2}(t)dz^{2}\,,
\end{equation}
where the mean scale-factor is $a=\sqrt[3]{ABC}$. In covariant terms, the only non-vanishing quantities in Bianchi I cosmologies are the relativistic energy density and pressure, the anisotropic stress tensor, the volume scalar, the shear and the electric Weyl tensor (i.e. $\rho$, $P$, $\pi_{ab}$, $\Theta$, $\sigma_{ab}$ and $E_{ab}$ respectively)~\cite{TCM}. All the remaining terms are zero by construction, namely $\omega_{a}=0=\dot{u}_{a}=q_{a}=H_{ab}=\mathcal{R}_{ab}$ ($\mathcal{R}_{ab}$ represents the 3-D counterpart of the Ricci tensor and it measures the curvature of the fundamental observers' rest-space). It is worth noting that because of their non-zero anisotropic stress tensor ($\pi_{ab}\neq 0$) Bianchi I models can generally host viscous fluids such as the electromagnetic ones, however under the restriction of zero momentum density ($q_{a}=0$). In case of an electromagnetic fluid the aforementioned limitation translates into a zero Poynting vector, $q^{\text{(em)}}_{a}=\epsilon_{abc}E^{b}B^{c}=0$, which means that on considering large-scale magnetic fields, the associated electric components of the Maxwell field have to vanish. This means that the Bianchi I cosmologies satisfy the MHD approximation by construction. Finally, we mention here for reference that the condition $\mathcal{R}_{ab}=0$ together with the continuity equation (for a Bianchi I model) are written as
\begin{equation}
    H^{2}=\frac{1}{3}(\rho+\frac{1}{2}\mathcal{B}^{2}+\sigma^{2})\hspace{15mm} \text{and} \hspace{15mm} \dot{\rho}=-3H(\rho+P)-\sigma^{ab}\pi_{ab}\,.
    \label{eqns: BianchiI}
\end{equation}
Note that the terms in the continuity equation do not include any contribution from the magnetic field. The above relations will be used in the following subsections.

\subsubsection{Evolution of the model}

The evolution of the magnetised Bianchi I model has been studied in detail and in various different works (e.g. see~\cite{Jac, RS}). However, we have not found anywhere yet an exact solution for the magnetic energy density coinciding with our own. Only an indirect verification of our result have we found in the literature, and it is mentioned below.

To begin with, in order to acquire some insight into the effects of the magnetic fields on the evolution of the cosmologies in question, let us assume that the anisotropy of the model is exclusively due to the presence of the magnetic field (i.e. matter is considered as a perfect fluid). Mathematically speaking this assumption means that the magnetic field has to be an eigenfunction of the shear tensor, namely
\begin{equation}
    \sigma_{ab}B^{b}=\xi B_{a}\,,
    \label{eqn:eigenvalue-B}
\end{equation}
where $\xi$ is the associated eigenvalue. Subsequently, on multiplying~\eqref{eqn:eigenvalue-B} by $B^{a}$ and defining $B^{a}\equiv \mathcal{B}n^{a}$, we determine the value of $\xi$ to be
\begin{equation}
    \sigma_{ab}B^{a}B^{b}=\Sigma \mathcal{B}^{2}=-\frac{1}{3}\Theta \mathcal{B}^{2}=\xi \mathcal{B}^{2}\rightarrow\xi=-\frac{\Theta}{3}\,.
    \label{eqn:eigenvalue-B2}
\end{equation}
It is remarkable that if we substitute our value of $\xi$ into equation (43) from reference~\cite{TM}, we restore relation~\eqref{magn-field-BianchiI} for the evolution of the magnetic field ($\xi=-\Theta/3$ corresponds to $\lambda=-\Theta/2$ and $\zeta=-1/2$ in~\eqref{magn-field-BianchiI}). This is an important, though indirect, verification of our result within the literature.
Besides, the magnetic field vector is in parallel an eigenfunction of the anisotropic magnetic stress tensor, $\pi^{\text{(M)}}_{ab}=-B_{a}B_{b}+(\mathcal{B}^{2}/3)h_{ab}$, so that
\begin{equation}
    \pi^{\text{(M)}}_{ab}B^{b}=-\frac{2}{3}\mathcal{B}^{2}B_{a}\,.
    \label{eqn:eigenvalue-PIB}
\end{equation}
Combining equations~\eqref{eqn:eigenvalue-B}--\eqref{eqn:eigenvalue-PIB} we arrive at
\begin{equation}
    \sigma_{ab}=\frac{\Theta}{2\mathcal{B}^{2}}\pi^{\text{(M)}}_{ab} \quad \text{and} \quad \sigma^{2}\equiv\frac{1}{2}\sigma_{ab}\sigma^{ab}=\frac{\Theta^{2}}{12}=\frac{3}{4}H^{2}\,.
    \label{eqn:sigma-squared}
\end{equation}
With the aid of eqs.~\eqref{eqns: BianchiI} and \eqref{eqn:sigma-squared} for a perfect and barotropic fluid ($P=w\rho$), $\rho\propto a^{-3(1+w)}$, we find out that the square of the shear and the scale factor evolve in accordance with
\begin{equation}
    \sigma^{2}=c_{1}a^{-3(1+w)}+c_{2}a^{-6}\hspace{15mm} \text{and} \hspace{15mm} H^{2}=c_{3}a^{-3(1+w)}+c_{4}a^{-6}
    \label{eqn:BianchiI-evol}
\end{equation}
respectively, where $c_{1}$, $c_{2}$ are constants. We observe that on the one hand, as the scale factor becomes large, the model approaches a FRW (with flat spatial sections) type of evolution, $a\propto t^{2/3(1+w)}$. On the other hand, as we approach the early stages of the universe, the model tends to a Kasner type of evolution, $\sigma\propto a^{-3}$ and $a\propto t^{1/3}$, which is characterised by the shear domination (e.g. see~\cite{TCM}). The aforementioned behavior at large and small-scales is in accordance with that of a non-magnetised Bianchi I cosmology with perfect fluid. Therefore, the difference between a magnetised and a non-magnetised model is theoretically found in their intermediate stages of evolution. In particular, equation~(\ref{eqn:BianchiI-evol}b) recasts into the solvable form
\begin{equation}
    \frac{{\rm d}a}{\rm{d}t}=\pm\sqrt{c_{1}a^{-1-3w}+c_{2}a^{-4}}\hspace{15mm} \text{or equivalently into} \hspace{15mm} c_{5}\frac{a^{2}{\rm d}a}{\sqrt{1+c_{6}a^{3(1-w)}}}=\rm{d}t\,,
    \label{eqn:eqn:BianchiI-evol2}
\end{equation}
where $c_{5}=c_{4}^{-1/2}$ and $c_{6}=c_{3}/c_{4}$ are constants. Let us solve the above equation for two characteristic values of the barotropic index $w$, namely $w=1/3$ (radiation) and $w=0$ (dust). Specifically, the integration of~\eqref{eqn:eqn:BianchiI-evol2} in the cases of radiation and dust\footnote{We consider the scale factor as a real quantity. In the former case ($w=1/3$), we make use of the substitution $a=c_{6}\tan{u}\rightarrow u = \arctan (a/c_{6})$ whilst in the latter case ($w=0$) of $u=1+c_{6}a^{3}$.} leads respectively to the solutions
\begin{equation}
    t=\mathcal{C}_{1}a\sqrt{a^{2}+\mathcal{C}_{2}}-\ln{\mid\sqrt{a^{2}+\mathcal{C}_{2}}+a}\mid+\mathcal{C}_{3} \hspace{15mm} \text{and} \hspace{15mm} a(t)=\sqrt[3]{\mathcal{C}_{4}t^{2}+\mathcal{C}_{5}t+\mathcal{C}_{6}}\,,
    \label{eqn:eqn:BianchiI-evol3}
\end{equation}
where $\mathcal{C}_{1},...,\mathcal{C}_{6}$ are constants. We observe that on large scales the square root term dominates in~(\ref{eqn:eqn:BianchiI-evol2}a) so that $a\propto t^{1/2}$, which is the evolution formula during the radiation era of the standard cosmological model (see also the following subsection). Moreover, the small--scale limit of~(\ref{eqn:eqn:BianchiI-evol2}b) leads to the above mentioned Kasner type solution $a\propto t^{1/3}$. On the other hand, approaching large--scales, the average scale factor increases with the cosmic time in accordance with $a\propto t^{2/3}$ (see \ref{eqn:eqn:BianchiI-evol3}b), which is the familiar evolution formula holding during the dust era of the standard cosmological model (refer to the following subsection).

\subsubsection{Magnetic density--radiation/dust equality}\label{ssec:FRW-Bianchi}

Let us close the unit regarding magnetic fields in cosmology by identifying the cosmic equality of magnetic energy density and radiation/dust in a magnetised Bianchi I model (filled with ideal fluid), and comparing it with its counterpart in a magnetised FRW model with flat spatial sections. In other words, we need to specify at which scales the ratios $\rho_{B}/\rho_{\text{rad}}$ and $\rho_{B}/\rho_{m}$ become equal to unity in magnetised Bianchi I models.

In the first place, let us consider a Friedmann background model with curved spatial sections. The isotropy and homogeneity of the model requires that all vector-tensor quantities (electromagnetic fields are included) as well as 3-D gradients vanish identically. Therefore, one has to study electromagnetic fields in perturbed FRW models (e.g. for a detailed approach see~\cite{T3}). Allowing for the presence of a weak electromagnetic field, we consider a linearly perturbed FRW model. Hence, to first order the equation of continuity~\eqref{continuity-eq} for radiation and dust is written as\footnote{Taking into account eq.~\eqref{eqn:el-field-prop} note that the electromagnetic term in~\eqref{continuity-eq} is of nonlinear order.}
\begin{equation}
    \dot{\rho}_{\text{rad}}=-4H\rho_{\text{rad}}\hspace{15mm} \text{and} \hspace{15mm} \dot{\rho}_{m}=-3H\rho_{m}
    \label{contin-radiation-and-dust}
\end{equation}
respectively, which are solved (recall that $H=\dot{a}/a$) to give the well known evolution formulae
\begin{equation}
    \rho_{\text{rad}}=\rho_{\text{rad}_{0}}\left(\frac{a_{0}}{a}\right)^{4}\hspace{15mm} \text{and} \hspace{15mm} \rho_{m}=\rho_{m_{0}}\left(\frac{a_{0}}{a}\right)^{3}\,,
    \label{rad-and-dust-densities}
\end{equation}
where the zero index corresponds to a specific cosmological instant. Moreover, assuming that the cosmic radiation is found in thermodynamic equilibrium, it can be approximated by the black--body radiation model. In particular, the radiation density has to be proportional to the fourth power of the cosmic fluid's absolute temperature $T$, in accordance with the Stefan-Boltzmann law
\begin{equation}
    \rho_{\text{rad}}=\sigma_{SB}T^{4}\,,
    \label{Stefan-Boltzmann}
\end{equation}
where $\sigma_{SB}=5.670\times 10^{-8}~\text{W}~\text{m}^{-2}~\text{K}^{-4}$ represents the Stefan-Boltzmann constant. Note that the combination of~(\ref{rad-and-dust-densities}a) and~\eqref{Stefan-Boltzmann} leads to the familiar relation $T\propto a^{-1}$, which is valid in both FRW and Bianchi I (with ideal fluid content) models. The radiation decays faster--due to the expansion of the universe--than the dust. These rates are expected to be modified in a Bianchi I model due to effects associated with the shear and vorticity. However, it can be easily checked that exactly the same relations for the density of radiation and dust hold in a Bianchi I model with ideal fluid content (recall that large--scale electric fields vanish by construction in a magnetised Bianchi I model). In this case, the geometric anisotropy comes exclusively from the large--scale magnetic fields. Regarding the magnetic energy density, it evolves according to the relations
\begin{equation}
    \rho^{\text{FRW}}_{B}=\rho^{\text{FRW}}_{B_{0}}\left(\frac{a_{0}}{a}\right)^{4}\hspace{15mm} \text{and} \hspace{15mm} \rho^{\text{Bianchi I}}_{B}=\rho^{\text{Bianchi I}}_{B_{0}}\left(\frac{a_{0}}{a}\right)^{6}\,,
    \label{magn-density-FRW-Bianchi}
\end{equation}
in a linearly perturbed FRW\footnote{The electric field density shares the same evolution formula with its magnetic counterpart--to first order with respect to a Friedmann background} with flat spatial sections and in an exact Bianchi I model respectively. It is worth noting that the radiation and the magnetic energy densities have the same rate of change in the former case, whereas this is not generally true in the latter case. In other words, although the electromagnetic field (or simply the magnetic field in the Bianchi I case) makes part of the radiation fluid, it does not necessarily evolve as the associated relativistic particles do.

Now taking into account the relations~\eqref{rad-and-dust-densities} and~\eqref{magn-density-FRW-Bianchi} we determine the ratio of the magnetic energy density over the density of radiation or dust, at a given moment in a Bianchi I model (with ideal fluid content) as\footnote{Note that $a$ represents now the average (with respect to all spatial directions) scale factor.}
\begin{equation}
     \frac{\rho_{B}}{\rho_{\text{rad}}}=\left(\frac{\rho_{B}}{\rho_{\text{rad}}}\right)_{p}\left(\frac{a_{p}}{a}\right)^{2}\hspace{15mm} \text{and} \hspace{15mm} \frac{\rho_{B}}{\rho_{m}}=\left(\frac{\rho_{B}}{\rho_{m}}\right)_{p}\left(\frac{a_{p}}{a}\right)^{3}\,,
     \label{densities-ratio}
 \end{equation}
 where the suffix $p$ indicates the values of the involved quantities at the present and $a_{p}/a=1+z$, with $z$ being the redshift. In accordance with the above expression, magnetic fields dominated in the past whilst their contribution to the total energy density is significantly limited today. When the two forms of energy acquire equal densities ($\rho_{\text{rad}}=\rho_{m}$), the corresponding scale factors ($a_{\text{eq}~(B-rad)}$ and $a_{\text{eq}~(B-m)}$) are
 \begin{equation}
     a_{\text{eq}~(B-rad)}=\left(\frac{\rho_{B}}{\rho_{\text{rad}}}\right)^{1/2}_{p}a_{p}\sim 10^{-9}a_{p}\hspace{15mm} \text{and} \hspace{15mm} a_{\text{eq}~(B-m)}=\left(\frac{\rho_{B}}{\rho_{m}}\right)^{1/3}_{p}a_{p}\sim 10^{-7.3}a_{p}\,,
     \label{aeq}
 \end{equation}
 namely about a billion and ten million times smaller respectively than today (the associated redshifts are $1+z_{\text{eq}~(B-rad)}=10^{9}$ and $1+z_{\text{eq}~(B-m)}=10^{7.3}$). In the above calculation we have taken into account that the present value of intergalactic magnetic fields amounts to the order of $10^{-15}$ Gauss (e.g.~refer to~\cite{Tav}--\cite{KKT}). Making use of natural units ($c=\hbar=k_{B}=1$) the intergalactic magnetic energy density today is expressed in terms of GeV's as $\rho_{B}\sim 4\times 10^{-70}~\text{GeV}^{4}$, in accordance with the equivalence: $1~ (\text{Gauss})^{2}/(8\pi)\simeq 2\times 10^{-40}~\text{GeV}^{4}$ (e.g.~see the appendix of~\cite{KolT}). Moreover, the density of matter today is $\rho_{m}\sim 10^{-30}~\text{gr}/\text{cm}^{3}\sim 4\times 10^{-48}~\text{GeV}^{4}$~($\rho_{m}=\Omega_{m}h^{2}\rho_{\text{crit}}$ with $\rho_{\text{crit}}\sim 10^{-29}~\text{gr}/\text{cm}^{3}$ and $\Omega_{m}h^{2}\simeq 0.14$ today~\cite{Planck}) whilst its radiation counterpart is $\rho_{\text{rad}}=10^{-34}~\text{gr}/\text{cm}^{3}\sim 4\times 10^{-52}~\text{GeV}^{4}$ ($1~\text{GeV}^{4}\simeq 2\times 10^{17}~\text{gr}/\text{cm}^{3}$). Moreover, with the aid of~(\ref{magn-density-FRW-Bianchi}b) and~\eqref{aeq} we calculate the values of the magnetic field at the aforementioned equalities and at recombination\footnote{Recombination takes place at redshift of about $1+z_{\text{rec}}=\frac{T_{\text{rec}}}{T_{p}}\simeq 1500$, where $T_{p}=2.7~\text{K}$ is the temperature of the Cosmic Microwave Background at present.} to be $B_{\text{eq}~(B-rad)}\sim 10^{12}$ G, $B_{\text{eq}~(B-m)}\sim 10^{7}$ G and $B_{\text{rec}}\sim 10^{-6}$ G respectively (the associated values of the densities are $4\times 10^{-16}~\text{GeV}^{4}$, $4\times 10^{-26.2}~\text{GeV}^{4}$ and $4\times 10^{-52}~\text{GeV}^{4}$).
 
 Before proceeding to a comparison of our results with their counterparts in a Friedmann model, let us raise and take into account an issue related to the constraint that cosmic nucleosynthesis imposes on the magnitude of the magnetic energy density. In particular, magnetic fields are known to increase nuclear reaction/transformation rates\footnote{Besides, magnetic fields contribute to the expansion rate of the universe and thus indirectly affect the rate of nuclear interactions.}, so that the enhanced domination of the magnetic energy density during the early Bianchi I universe ($\rho_{B}\propto a^{-6}$ instead of $\rho_{B}\propto a^{-4}$ in a Friedmann model) may potentially be incompatible with the cosmic nucleosynthesis. We attempt here a first approach to the question by comparing the densities of magnetic fields and radiation during nucleosynthesis. In practice, considering that nuclear binding energies are of the order of some MeV, which correspond (in thermal--statistical equilibrium) to absolute temperatures of the order $T_{\text{NS}}\sim 1~\text{MeV}/(k_{B}=8.61\times 10^{-11}~\text{MeV~K}^{-1})\sim 10^{10}$~K ($k_{B}$ is the Boltzmann constant), we can estimate that nucleosynthesis within the standard cosmological model takes place at redshift
 \begin{equation}
     1+z_{\text{NS}}=\frac{T_{\text{NS}}}{T_{p}}\sim 10^{9}\,,\hspace{15mm} \text{which means that} \hspace{15mm} a_{\text{NS}}\sim 10^{-9}a_{p}\,.
     \label{eqn:nucleos}
 \end{equation}
 It is straightforward to observe (comparing~(\ref{aeq}a) and~(\ref{eqn:nucleos}b)) that in the context of a Bianchi I model (with perfect fluid content), magnetic fields and radiation share approximately (an order of magnitude estimation) the same densities during nucleosynthesis. At a first glance, the small difference we find in densities seems not to permit us to derive any conclusion. However, we shall keep in mind that our estimation depends on the value, which we have assumed, of the intergalactic magnetic field today (i.e. $B_{p}\sim 10^{-15}~G$). For instance, a weaker magnetic field, such as $B_{p}\sim 10^{-16}~G$, can lead to a ratio $(\rho_{B}/\rho_{\text{rad}})_{\text{NS}}\sim 10^{-2}$, which shows a clear domination of radiation over magnetic fields during the epoch of nucleosynthesis. Such a significant difference (of two orders of magnitude) seems to favor the answer that the presence of magnetic fields does not disturb the cosmic creation of nuclei.
 
 Now in analogy with relation~\eqref{aeq}, the equality of magnetic energy density and dust in a perturbed (magnetised) Friedmann model with flat spatial sections takes place at $a_{\text{eq}~(B-m)}\sim 10^{-22}a_{p}$ (or equivalently at $1+z_{\text{eq}~(B-m)}\sim 10^{22}$), namely at a redshift about fifteen orders of magnitude greater than its Bianchi I counterpart. This means that in a Bianchi I cosmology the magnetic energy density of the highly conducting cosmic fluid is overwhelmed by the energy density of dust much later during the universe's evolution in comparison to a Friedmann model. As for the ratio $\rho_{B}/\rho_{\text{rad}}$, it remains constant throughout the evolution of the magnetised FRW model, because magnetic fields and radiation share the same expansion rate. On the other hand, the equality of magnetic fields with dust occurs after their equality with radiation, while both equalities take place much earlier (during the radiation era) than the recombination as well as than the dust-radiation equality. The aforementioned results could hopefully turn out to be useful when examining the potential cosmological origin of magnetic fields in the pre--recombination epoch.

\section{Gravitational collapse of a magnetised fluid}
\label{sec:Grav-collapse}

The gravitational collapse of compact stellar objects, like white dwarfs, neutron stars, black holes, as well as that of protogalactic clouds usually involves (weak or strong) magnetic fields. In the context of general relativity, independent studies have pointed out the unconventional tendency of the $B$-fields  to resist their own gravitational implosion. The same works have also raised the question as to whether the magnetic presence and the resulting Lorentz forces could actually halt the contraction of the surrounding collapsing matter~\cite{Mel}--\cite{TMav}. In addition, alternative studies of charged collapse, this time employing the repulsive (electrostatic) Coulomb forces, have found that the latter could also prevent the formation of spacetime singularities~\cite{N}--\cite{R}. The present section probes the gravitational collapse of a highly conductive charged medium by means of the Raychaudhauri equation and along the lines of~\cite{T3}--\cite{TMav}. Making a step further, we take advantage of a 1+2 spatial splitting and arrive at a simple criterion which could decide the ultimate fate of homogeneously contracting magnetised media. This criterion is then applied to a collapsing perturbed Bianchi I spacetime permeated by a magnetic field.

\subsection{Using the Raychaudhuri equation}\label{ssec:Raychaudhuri}

Traditionally, theoretical studies of gravitational collapse make use of the Raychaudhuri equation which has been made famous as a keystone of the singularity theorems. Besides, in general terms, the formula in question covariantly describes the volume evolution of a self--gravitating fluid element. In this first subsection, we revisit the problem of gravitational implosion of a highly conducting (magnetised) fluid with the aid of the Raychaudhuri equation\footnote{Apart from its conventional application to timelike worldlines of real (or hypothetical) observers, the aforementioned equation has been applied to spacelike and null curves as well (e.g.~see~\cite{KS,AV}).}, and in light of our new knowledge regarding the behavior of the associated magnetic field (more specifically of relation~\eqref{eqn:bexp}), as well as of our new developments in the context of the 1+1+2 covariant formalism. Unlike previous independent works, our study builds upon past research (see~\cite{T3}-\cite{TMav}) and leads to a remarkably simple criterion determining the fate of homogeneous and magnetised gravitational collapse.

Before proceeding to the analysis, let us have in mind two crucial points. Firstly, magnetic-line deformations are usually caused by electrically charged particles, however relativistic spacetime curvature (gravity) also potentially behaves as a deforming agent~\cite{T2, TMav}. Secondly, the magnetic tension reflects the elasticity of the field lines and their tendency to react against any agent that distorts them from equilibrium~\cite{T3, KT, TMav}.

Let us start with the Raychaudhuri equation, which we have already written in the form of~\eqref{Raych-eqn1}. To proceed, we need to calculate the 3-divergence of the acceleration vector (i.e. ${\rm D}^{a}\dot{u}_{a}$) which gives rise to magneto-geometric terms, of crucial importance for our relativistic study. 
In particular, let us consider an ideal, highly conducting fluid model. Euler's equation is written thus as
\begin{equation}
    (\rho+P+B^{2})\dot{u}_{a}=-{\rm D}_{a}P-\frac{1}{2}{\rm D}_{a}B^{2}+B^{b}{\rm D}_{b}B_{a}+\dot{u}^{b}B_{b}B_{a}\,,
    \label{Euler's-MHD}
\end{equation}
where contributions from both matter and magnetic fields appear on its right-hand side. In order to facilitate the analytic calculations, we assume that the contracting fluid has nearly homogeneous matter\footnote{Note that the homogeneity of the matter fields is a rather common approximation. In fact, spatial homogeneity is a standard assumption in all typical singularity theorems~\cite{W, HE}. Besides, the assumption of homogeneous matter distribution does not essentially affect the validity of our argument, since gradients in the fluid and in the magnetic density distribution tend to inhibit gravitational contraction, even within Newtonian physics.} and magnetic energy density distributions (${\rm D}_{a}\rho\simeq 0\simeq {\rm D}_{a}P\simeq {\rm D}_{a}B^{2}$, where a barotropic equation of state, $P=w\rho$ with $w=\text{const}$, has been considered). However, we allow for $B^{b}{\rm D}_{b}B_{a}\neq 0$, so that we can study effects caused by distortions of the magnetic forcelines (see the following discussion). Subsequently, taking the 3-divergence of~\eqref{Euler's-MHD} in combination with the 3-Ricci identities (eq.~\eqref{3-D-Ricci}) and Maxwell's equations (eq.~\eqref{magn-div}) we arrive at
\begin{equation}
    {\rm D}^{a}\dot{u}_{a}=c^{2}_{\mathcal{A}}\mathcal{R}_{ab}n^{a}n^{b}+2(\sigma^{2}_{B}-\omega^{2}_{B})\,,
    \label{3-div-of-dot-u}
\end{equation}
where the scalars $\sigma^{2}_{B}={\rm D}_{\langle b}B_{a\rangle}{\rm D}^{\langle b}B^{a\rangle}/2(\rho+P+B^{2})$ and ${\rm D}_{[b}B_{a]}{\rm D}^{[b}B^{a]}/2(\rho+P+B^{2})$ represent the magnetic analogues of the shear and the vorticity respectively. Of special interest is the purely relativistic (magneto-geometric) term $\mathcal{R}_{ab}n^{a}n^{b}$ which describes 3-D distortions of the magnetic forcelines due to the curvature of the host spacetime. Note that all the terms on the right-hand side of ~\eqref{3-div-of-dot-u} are tension stresses triggered by the deformation of the magnetic field lines. Each of these terms acts against the agent that caused the deformation in the first place (e.g. the magneto-vorticity $\omega^{2}_{B}$ is caused by rotational effects, $\omega^{2}$, and it tends to counterbalance them. Observe the opposite signs of the pairs $\omega^{2}$, $\omega^{2}_{B}$ and $\sigma^{2}$, $\sigma^{2}_{B}$ in~\eqref{Raych-eq2}). Substituting expression~\eqref{3-div-of-dot-u} into the Raychaudhuri equation~\eqref{Raych-eqn1}, the latter reads
\begin{equation}
    \dot\Theta+\frac{1}{3}\Theta^{2}=-R_{ab}u^{a}u^{b}+c^{2}_{\mathcal{A}}\mathcal{R}_{ab}n^{a}n^{b}-2(\sigma^{2}-\sigma^{2}_{B})+2(\omega^{2}-\omega^{2}_{B})+\dot{u}^{a}\dot{u}_{a}\,,
    \label{Raych-eq2}
\end{equation}
where $R_{ab}u^{a}u^{b}=(\rho+3P+B^{2})>0$ represents the total (gravitational) energy density of the system. Note that if $\dot\Theta+\frac{1}{3}\Theta^{2}<0$, the above equation implies that an initially contracting congruence of worldlines will focus at a point ($\Theta\rightarrow -\infty$) within finite proper time. Hence, positive terms on the right-hand side of the Raychaudhuri formula act against the gravitational collapse whilst negative ones in the inverse way.

Having in mind the strong gravity conditions which characterise collapsing compact stellar objects (and the counterbalancing relation of the paired terms in~\eqref{Raych-eq2}), we choose to focus our attention on the purely relativistic--curvature terms\footnote{Note that $\dot{u}^{a}\dot{u}_{a}>0$ always, and therefore it resists contraction in any case.} (i.e. $c^{2}_{\mathcal{A}}\mathcal{R}_{ab}n^{a}n^{b}$ which is positive in all cases of realistic gravitational collapse and thus tends to inhibit the gravitational pull of the local matter, as encoded in the expression $R_{ab}u^{a}u^{b}$). Regarding the magneto-geometric tension stress $c^{2}_{\mathcal{A}}\mathcal{R}_{ab}n^{a}n^{b}$, it is expected to grow strong with increasing curvature distortion during the collapse, in analogy with the resisting power of a compressed elastic medium. In particular, if at some time during the implosion the following condition holds
\begin{equation}
    c^{2}_{\mathcal{A}}\mathcal{R}_{ab}n^{a}n^{b}>R_{ab}u^{a}u^{b}\,,
    \label{Grav-coll-cond}
\end{equation}
we expect that the latter will be halted.  Making use of the Gauss--Codacci formula (e.g. see expression (1.3.39) in~\cite{TCM}), the above condition transforms into
\begin{equation}
    2c^{2}_{\mathcal{A}}(\rho-\frac{1}{3}\Theta^{2})+3c^{2}_{\mathcal{A}}(E_{ab}-\frac{1}{3}\Theta\sigma_{ab}+\sigma_{ca}\sigma^{c}{}_{b}-\omega_{ca}\omega^{c}{}_{b})n^{a}n^{b}>\frac{3}{2}(\rho+3w\rho+\mathcal{B}^{2})\,,
\end{equation}
where the first of the two parentheses in the left-hand side represents the isotropic part of the tension stress whilst the second the anisotropic. It turns out that the latter must be nonzero which implies that the gravitational collapse has to be anisotropic, if the tension stress is to outbalance the gravitational pull of the matter.

\subsection{A non--collapse criterion}\label{ssec:1+2-split-collapse}

Once again we can take advantage of a 1+2 spatial split as well as of our newly gained knowledge regarding the evolution of the magnetic and the matter density fields (at the ideal MHD limit), to acquire physical insight into our problem. In particular, taking into account that $\mathcal{E}\equiv E_{ab}n^{a}n^{b}$, $\Sigma\equiv \sigma_{ab}n^{a}n^{b}=-\Theta/3$, $\sigma_{ca}\sigma^{c}{}_{b}n^{a}n^{b}=\Sigma^{2}+\Sigma^{a}\Sigma_{a}=\frac{1}{9}\Theta^{2}+\Omega^{a}\Omega_{a}$ (refer to expressions~\eqref{Sigma-def} and~\eqref{Sigma-vec-def} in the Appendix), $\omega_{ca}\omega^{c}{}_{b}n^{a}n^{b}=\Omega^{a}\Omega_{a}$ and the definition of the Alfv\'{e}n speed, our condition simplifies subsequently to
\begin{equation}
    (2\rho+3\mathcal{E})c^{2}_{\mathcal{A}}>\frac{3}{2}(\rho+3w\rho+\mathcal{B}^{2})
\end{equation}
and
\begin{equation}
    \mathcal{E}>\frac{1}{2}(1+4w+3w^{2})\left(\frac{\rho}{\mathcal{B}}\right)^{2}+\frac{1}{3}(1+6w)\rho+\frac{1}{2}\mathcal{B}^{2}\,.
    \label{coll-cond2}
\end{equation}
It is worth noting that the effects of rotation, associated with $\Omega^{a}\Omega_{a}$, and included in the term $\mathcal{R}_{ab}n^{a}n^{b}$, exactly cancel out. This happens because (in parallel it means that) the 3-D curvature deformation of the magnetic field-lines along their own direction is not affected by rotations (In particular, rotations of the surface shaped by the magnetic field direction and $\Omega^{a}$-for the case in question.). Now recall that the continuity equation for our fluid model (refer to~\eqref{continuity-eq3}), accepts solution~\eqref{dens-evol2}. According to the latter, the density of matter increases with a rate generally smaller than that for the magnetic energy density (i.e. $1+w\leq 2$). Especially in the case of stiff matter ($w=1$), the two growing rates are the same.

Allowing sufficient time for the collapse to evolve, we expect (considering relations~\eqref{eqn:bexp} and~\eqref{dens-evol2}) that the dominant term in~\eqref{coll-cond2} will be $\mathcal{B}^{2}$, so that
\begin{equation}
    \mathcal{E}>\frac{1}{2}\mathcal{B}^{2}\,.
    \label{implosion-criterion}
\end{equation}
In other words, if at some time during the collapse, the electric Weyl tensor along the magnetic forcelines prevails over the magnetic energy density, the collapse will turn into expansion and the system will be prevented from reaching a singularity\footnote{The following issue should be kept in mind when dealing with the problem of magnetised gravitational implosion. Under their continuous and increasing deformation during the collapse (due to the increasing spacetime curvature), the magnetic forcelines may lose their elastic properties and ultimately be broken. Hence, the questions raised by such a possibility could be the object of a potential research work in the future. In particular, \textit{what happens with the magnetic fieldlines at an advanced stage of the collapse? Will they inevitably be broken and when? Will they reconnect? Can they definitely affect or specify the fate of the collapse before having lost their elasticity or before being broken?}}. More specifically, recall that on the one hand $\mathcal{E}$ encodes the tidal forces acting upon the magnetic field-lines and resisting to their spatial distortion (see also the discussion regarding the term $c^{2}_{\mathcal{A}}\mathcal{R}_{ab}$ in the previous subsection). These (increasing in value) forces are triggered by the geometric deformation of the magnetic field-lines due to the increasing gravitational energy density of the system ($-R_{ab}u^{a}u^{b}$) during the contraction. The agent responsible for the resistance of the magnetic forcelines to their deformation, and consequently for the creation and reinforcement of $\mathcal{E}$, is the tension stress associated with their elasticity. On the other hand side of~\eqref{implosion-criterion}, the (increasing according to~\eqref{eqn:bexp}) magnetic energy density $\rho_{B}=\mathcal{B}^{2}/2$ acts in the opposite way by contributing to the total gravitational mass-energy of the system and thus enhancing the collapse process. To illustrate further our criterion, let us recall that in terms of Newtonian gravity, $E_{ab}$ is associated with the second-order derivative of the gravitational potential $\Phi$ (precisely the Newtonian tidal tensor) or equivalently with the first-order derivative of the tidal forces $F$, in accordance with (e.g. see~\cite{EMM})\footnote{In the context of Newtonian theory, studying tidal forces presupposes the consideration of at least two distinctive massive bodies. However, from a relativistic point of view, we can envisage tidal forces as a result of the different curvature effects (caused by the fluid's spacetime energy distribution) experienced by distinctive particles of the magnetised fluid.}
\begin{equation}
    E^{\text{(Newt)}}_{ab}=\partial_{a}\partial_{b}\Phi-\frac{1}{3}(\partial^{c}\partial_{c}\Phi) h_{ab}\hspace{15mm} \text{and} \hspace{15mm} \mathcal{E}^{\text{(Newt)}}=\mathcal{F}'-F^{a}n'_{a}\,,
    \label{Weyl-Newtonian}
\end{equation}
where the latter relation comes from the double projection of the former along $n^{a}$ and $\mathcal{F}=F^{a}n_{a}$, $F^{a}n'_{a}$ correspond to tidal forces acting along and normal to the magnetic forcelines respectively.

Predicting actually the fate of almost homogeneous gravitational collapse of a highly conducting fluid remains an open question. Our results indicate that the latter question reduces to whether the electric Weyl tensor along the magnetic field lines increases faster than the magnetic energy density or not. The answer seems to depend on the geometric background in hand, and potentially on the problem's initial conditions.

\subsection{Studying magnetised collapse on a perturbed Bianchi I background}

In order to put in practice our criterion for the gravitational implosion of a magnetised fluid~\eqref{implosion-criterion}, we need to adopt a specific geometric model. In the first place, an appropriate model has to satisfy three principal requirements; on the one hand, to be by construction homogeneous and a natural host of pure, large-scale magnetic fields\footnote{The latter requirement implies that the model has to be anisotropic as well. In fact, two simple and familiar models within astrophysics and cosmology, namely the Schwarzschild and the Friedmann-Robertson-Walker geometries, could not be appropriate candidates for our analysis, due to the aforementioned requirements}; on the other hand, to have closed spatial sections and be contracting, if we want to establish a correspondence between our model and the collapse of a stellar object or a protogalactic cloud. In case we adopt a model at the perturbation level, the first two restrictions have to be satisfied in the background geometry. This is necessary, regarding the latter, because our relation for the evolution of the magnetic field holds exactly at the MHD limit. Concerning the former, the homogeneity of the background is needed in practice for considering gauge-invariant perturbations (quantities which remain constant or vanish in the background) in accordance with the Stewart-Walker lemma~\cite{SW} (see the analysis below). As for the third requirement, we do not have a specific reason for demanding its satisfaction in the background. Overall, our choice seems to be directed-at least by the first two requirements-towards the family of the homogeneous and anisotropic Bianchi models, some of which (namely I, II, III, $\text{VI}_{-1}$ and $\text{VII}_{0}$) can accommodate constrained magnetic field components~\cite{HJ}. Now of the Bianchi spacetimes only IX is known to have positive curvature geometry (e.g. see~\cite{EMM}). Therefore, none of the Bianchi models seems appropriate to describe exactly the phenomenon of homogeneous and magnetised gravitational collapse. The simplest available choice coming into view is to study the Bianchi I model (with Euclidean spatial geometry) at the linear perturbation level--which allows us to construct closed geometric sections.

More specifically, in what follows we consider the propagation of the electric Weyl tensor in reference to a (magnetised) Bianchi I type geometric background. The basic geometric-dynamic and kinematic quantities describing a Bianchi I spacetime have been outlined in subsection~\ref{ssec:BianchiI}. To proceed, we need to consider the 3-D Ricci tensor $\mathcal{R}_{ab}$ (consequently the spatial gradients of the magnetic field as well--see subsection~\ref{ssec:Raychaudhuri}) and the 4-acceleration $\dot{u}_{a}$ (recall eq.~\eqref{3-div-of-dot-u} and the associated analysis) as first-order perturbations\footnote{The magnetic Weyl tensor $H_{ab}$ is also a perturbation not appearing at present. See~\eqref{dot-Eab} in the following, where it makes its first appearance.} in reference to our background. Repeating the reasoning--which remains exactly the same--described in subsections~\ref{ssec:Raychaudhuri} and~\ref{ssec:1+2-split-collapse}, it is straightforward to conclude that the collapse criterion~\eqref{implosion-criterion} holds in our linearly perturbed Bianchi I model. Moreover, we ensure that the model has closed spatial sections by imposing the positive sign condition of the 3-D Ricci tensor $\mathcal{R}_{ab}$ along every spatial direction. Specifically, along the magnetic field lines (see relation~\eqref{coll-cond2}) and regarding the 3-D Ricci scalar (e.g. refer to eq. (1.3.40) in~\cite{TCM}), the aforementioned condition takes the form
\begin{equation}
   \mathcal{R}_{ab}n^{a}n^{b}=2\rho+3\mathcal{E}>0\Rightarrow\mathcal{E}>-\frac{2}{3}\rho \hspace{15mm} \text{and} \hspace{15mm} \mathcal{R}=2(\rho-\frac{1}{3}\Theta^{2}+\sigma^{2})>0
    \label{Ricci-condition}
\end{equation}
respectively. Of particular interest is the former which sets a lower boundary of $\mathcal{E}$ (given that $\rho>0$). Subsequently, aiming to focus on the evolution of the electric Weyl curvature tensor $E_{ab}$ we shall firstly have a look at its general propagation equation, which is (e.g. see~\cite{TCM})
\begin{eqnarray}
    \dot{E}_{\langle ab\rangle}&=&-\Theta E_{ab}-\frac{1}{2}(\rho+P)\sigma_{ab}+\text{curl}H_{ab}-\frac{1}{2}\dot{\pi}_{ab}-\frac{1}{6}\Theta\pi_{ab}-\frac{1}{2}{\rm D}_{\langle a}q_{b\rangle}-\dot{u}_{\langle a}q_{b\rangle} \nonumber\\
    &&+3\sigma_{\langle a}{}^{c}\left(E_{b\rangle c}-\frac{1}{6}\pi_{b\rangle c}\right)+\epsilon_{cd\langle a}\left[2\dot{u}^{c}H_{b\rangle}{}^{d}-\omega^{c}\left(E_{b\rangle}{}^{d}+\frac{1}{2}\pi_{b\rangle}{}^{d} \right) \right]\,.
    \label{dot-Eab}
\end{eqnarray}
Under our homogeneity and perfect fluid assumptions the linearisation of the above equation (at the MHD limit) with respect to the Bianchi I background leads to
\begin{equation}
\dot{E}_{\langle ab\rangle}=-\Theta E_{ab}-\frac{1}{2}(\rho+P)\sigma_{ab}-\frac{1}{2}\dot{\pi}_{ab}-\frac{1}{6}\Theta\pi_{ab}+3\sigma_{\langle a}{}^{c}\left(E_{b\rangle c}-\frac{1}{6}\pi_{b\rangle c}\right) \nonumber\,,
    \label{dot-Eab-2}
 \end{equation}
where the anisotropic pressure input comes from the magnetic field only (recall that $\pi^{\text{(magn)}}_{ab}=-\mathcal{B}^{2}n_{\langle a}n_{b\rangle}$--see subsection~\ref{subsec:Fluid-des}). Moreover, note that the assumption of homogeneity imposes that the magnetic Weyl component $H_{ab}$ vanishes at the linear level (e.g. refer to eq. (1.3.8) in~\cite{TCM}). Subsequently, as we are interested in the evolution of $\mathcal{E}\equiv E_{ab}n^{a}n^{b}$, we project relation~\eqref{dot-Eab-2} along $n^{a}$ (with respect to both indices), so that it finally transforms into
\begin{equation}
    \dot{\mathcal{E}}+\frac{5}{2}\Theta\mathcal{E}-\frac{1}{6}(1+w)\Theta\rho+\frac{1}{2}\Theta \mathcal{B}^{2}=0\,.
    \label{prop-of-E-cal}
\end{equation}
The above\footnote{Note that it consists of a gauge--invariant equation, where no quantity represents a perturbation.} is a linear, partial differential equation (note that $\mathcal{E}$ presents spatial dependence) of first-order. In order to proceed to its solution, we adopt a frame parallelly propagated along the worldlines (or the collapsing fluid), so that $\dot{\mathcal{E}}={\rm d}\mathcal{E}/{\rm d}\tau=\partial\mathcal{E}/\partial\tau+(\partial_{i}\mathcal{E})u^{i}$, where the last term vanishes by making use of comoving coordinates. On taking into account expressions~\eqref{eqn:bexp} and~\eqref{dens-evol2}, equation~\eqref{prop-of-E-cal} is solved in the standard way giving
\begin{equation}
    \mathcal{E}=\mathfrak{B}e^{-\frac{5}{2}\int{\Theta_{0}d\tau}}-\mathcal{C}^{2}e^{-2\int{\Theta_{0}d\tau}}+\left(\frac{1+w}{9-6w}\right)\mathcal{D}e^{-(1+w)\int{\Theta_{0}d\tau}}\,,
    \label{E-cal-sol}
\end{equation}
where $\mathfrak{B}$, $\mathcal{C}$ and $\mathcal{D}$ are constants (see equations~\eqref{eqn:bexp} and~\eqref{dens-evol2}). Note that the above relation describes the temporal evolution of $\mathcal{E}$ with respect to proper time $\tau$ (i.e. the parameter of the worldlines). During the implosion, ($\Theta_{0}<0$) the electric Weyl curvature along the magnetic forcelines increases (under the assumption of continuity, so that $\int{\Theta_{0}d\tau}<0$) according to three different terms, which correspond to the contributions of the magnetic and matter energy densities, as well as of the term $(5\Theta\mathcal{E})/2$ in the left-hand side of~\eqref{E-cal-sol}. The maximum variation of $\mathcal{E}$ comes from the exponential term with coefficient two (recall that the maximum value of $1+w$ is two as well, $w\leq 1$), which means that it does not increase faster than $\mathcal{B}^{2}$. Therefore, it seems that the fate of our collapse model-whether criterion~\eqref{implosion-criterion} is satisfied or not- basically depends on the problem's initial conditions. 

\section{Discussion}

On decomposing Faraday's equation into its 1 temporal and 1+2 spatial components, we have shown that it can be solved independently at the MHD limit leading to a solution for the magnetic field. In particular, we have found out that the magnetic energy density generally increases or decreases in accordance with the inverse cube law of the scale factor (a result which is--as far as we know--new in the literature) associated with the fluid's continuous contraction or expansion respectively. Alternatively, this type of change corresponds to an exponential spacetime function with a negative integral of the volume scalar (actually of its individual components) in its exponent. An analogous relation holds for the matter density of an ideal fluid. The aforementioned solutions in combination with Euler's equations of motion, the continuity equation, an equation of state and Raychaudhuri equation, provide a description of the magnetic field's behavior in relation to the motion of the self-gravitating, highly conducting fluid. More specifically, we have pointed out that the magnetic force terms tend to dominate over the pressure or matter density gradient in the case of contraction ($\Theta<0$), determining thus the quantity and the direction of the fluid's motion. Inversely, the domination of matter is expected to take place in the case of expansion ($\Theta>0$). Besides, we have noted the aforementioned conclusion holds under the assumption that the evolution of the volume scalar $\Theta$ is of minor importance in comparison to that of $\rho$ and $\mathcal{B}^{2}$. 

When applied to homogeneous and anisotropic (magnetised) cosmological models, relation~\eqref{eqn:bexp} tells us that the magnetic energy density--hence the total radiation density in the MHD limit--is proportional to the inverse sixth power of the mean, time dependent scale factor. Especially regarding a Bianchi I model, consisting of a magnetised perfect fluid, our field's law of variation finds a remarkable, indirect verification within the literature. Moreover, on deriving the evolution formulae of the model in question (see eqs~(\ref{eqn:eqn:BianchiI-evol3}a,~b)), we have found out that they reduce to the standard cosmic radiation and dust expansion/contraction formulae at the small and the large--scale limit respectively. Another remarkable result is that as a consequence of the significant difference in the rate of change of the magnetic energy density between a magnetised Bianchi I and a perturbed FRW model, the epoch of magnetic energy and matter densities equality in the former case corresponds to a redshift which is about fifteen orders of magnitude smaller than its counterpart in the latter case. This difference should probably be taken into account when searching for the origin of cosmic magnetic fields during the pre-recombination era. Overall, large--scale magnetic fields are known to constitute a real component of the universe and thus contribute to its total energy content. Therefore, the knowledge of their evolution formula can provide a valuable tool when dealing with the dynamics of realistic cosmological models.

We have also examined a crucial astrophysical application of relation~\eqref{eqn:bexp}, namely the gravitational collapse of a magnetised fluid. In particular, studying the contracting worldlines with the aid of the Raychaudhuri formula, we conclude that if at some time during the homogeneous (in reference to matter and magnetic energy densities) implosion, the electric Weyl tensor along the magnetic forcelines overwhelms the magnetic energy density, then the gravitational contraction will be prevented from reaching a singularity. Our result gives rise to the following question: which of the two rivalling terms, the electric Weyl curvature and the magnetic energy density, increases faster, so that it finally dominates? Given that the way $\mathcal{B}^{2}$ changes, is known, the above question reduces to determining the evolution of $\mathcal{E}$. The answer seems to depend on the geometric background one adopts. Making a step towards testing our implosion criterion, we have adopted an homogeneous, linearly perturbed (so that it approximately has closed spatial sections) Bianchi I model  of magnetised collapse. Our results show that the electric Weyl curvature can not increase faster than the magnetic energy density for the model in question. As a consequence, the fate of the collapse seems to be in principle a matter of initial conditions. Our implosion model has the advantage of not being restricted by many assumptions (basically homogeneity and perfect fluid energy content, which are standard), while perturbations are needed only to construct closed spatial geometry. Nevertheless, it would definitely be better if one found an exact\footnote{Recall that for an analytic approach we have searched for a homogeneous model, natural host of pure, large-scale magnetic fields, with closed spatial sections.} (unperturbed) model for studying the collapse of a highly conducting fluid.

The magnetic field's law of variation for a highly conducting fluid--our fundamental conclusion--is expected to provide magnetohydrodynamics with a valuable new theoretical tool. Overall, our results could hopefully, on the one hand, shed new light on the description of magnetised compact stellar objects such as black holes, neutron stars (of particular interest are pulsars and magnetars) and white dwarfs. In parallel, a verification of our results could be given by studies of the aforementioned objects. On the other hand, in reference to the field of cosmology, our exact (not approximate) evolution formula for the magnetic field could fortunately refresh the question concerning the energy contribution of large--scale magnetic fields to the kinematics of our universe.

\appendix

\section{The physical content of the 1+2 components of the shear}\label{Shear-1+2-comp}

In what follows, we reveal some relations between the 1+2 components of the shear and other kinematic quantities. These relations are of great importance when dealing with the split calculations in sections~\ref{sec:1+2-split},~\ref{sec:Cosmol-magn-fields} and~\ref{sec:Grav-collapse}.

To begin with, let us consider the definition of $\Sigma$ and simply follow the operations

\begin{equation}
\Sigma\equiv\sigma_{ab}n^{a}n^{b}\equiv {\rm D}_{\langle b}u_{a\rangle}n^{a}n^{b}={\rm D}_{(b}u_{a)}n^{a}n^{b}-\frac{1}{3}\Theta h_{ab}n^{a}n^{b}=u'_{a}n^{a}-\frac{1}{3}\Theta\,.
\end{equation}
Therefore, $\Sigma$ is a quantity which expresses the fluid's volume expansion/contraction according to the relation
\begin{equation}
\Sigma=-\frac{1}{3}\Theta\,.
\label{Sigma-def}
\end{equation}
In the same way, by the definition of $\Sigma_{a}$ we have
\begin{equation}
    \Sigma_{a}\equiv \tilde{h}_{a}{}^{b}\sigma_{bc}n^{c}\equiv \tilde{h}_{a}{}^{b}n^{c}{\rm D}_{\langle c}u_{b\rangle}=\tilde{h}_{a}{}^{b}n^{c}{\rm D}_{(c}u_{b)}-\frac{1}{3}\tilde{h}_{a}{}^{b}n^{c}(\Theta h_{cb})=\frac{1}{2}\tilde{h}_{a}{}^{b}u'_{b}\,.
    \end{equation}
Therefore, $\Sigma_{a}$ is a quantity equivalent to the derivative of the 4-velocity along the vector $n^{a}$ according to the relation
\begin{equation}
\Sigma_{a}=\frac{1}{2}u'_{a}\,.
\end{equation}
Furthermore, consider now the expression $({\rm D}_{a}u_{b})n^{b}$-which is equal to $-({\rm D}_{a}n^{b})u_{b}=0$ in accordance with Leibniz's rule- and decompose the spatial derivative of the 4-velocity
\begin{equation}
({\rm D}_{a}u_{b})n^{b}=(\sigma_{ab}-\omega_{ab}+\frac{1}{3}\Theta h_{ab})n^{b}=\Sigma n_{a}+\Sigma_{a}+\epsilon_{ac}\Omega^{c}+\frac{1}{3}\Theta n_{a}=0\,.
\end{equation}
Projecting orthogonal to $n^{a}$ the above equation becomes
\begin{equation}
\Sigma_{a}=-\epsilon_{ab}\Omega^{b}\,,
\label{Sigma-vec-def}
\end{equation}
which means that $\Sigma_{a}$ is a vector almost equivalent to the vorticity vector $\Omega^{a}$ (note that the two vectors are orthogonal to each other and have the same length), both lying on the 2-surface normal to $n^{a}$.

Finally, starting from the definition of $\Sigma_{ab}$ we have
\begin{equation}
    \Sigma_{ab}\equiv \left(\tilde{h}_{(a}{}^{c}\tilde{h}_{b)}{}^{d}-\frac{1}{2}\tilde{h}_{ab}\tilde{h}^{cd}\right)\sigma_{cd}=\frac{1}{2}\tilde{h}_{a}{}^{c}\tilde{h}_{b}{}^{d}{\rm D}_{(d}u_{c)}+=\frac{1}{2}\tilde{h}_{b}{}^{c}\tilde{h}_{a}{}^{d}{\rm D}_{(d}u_{c)}-\frac{1}{2}\tilde{h}_{ab}\tilde{h}^{cd}{\rm D}_{(d}u_{c)}\,,
\end{equation}
which becomes
\begin{equation}
    \Sigma_{ab}={\rm D}_{(b}u_{a)}-n_{(b}u'_{a)}-\frac{1}{2}\Theta\tilde{h}_{ab}=\tilde{{\rm D}}_{(b}u_{a)}-\frac{1}{2}\Theta\tilde{h}_{ab}\,.
\end{equation}
Consequently, we find out that
\begin{equation}
    \Sigma_{ab}=\tilde{{\rm D}}_{\langle b}u_{a\rangle}\,,
    \label{def-for-Sab}
\end{equation}
namely that $\Sigma_{ab}$ consists of the 2-D counterpart of the 3-dimensional gradient of the 4-velocity field--recall that $\sigma_{ab}\equiv{\rm D}_{\langle b}u_{a\rangle}$.

\section{1+2 decomposition of the full Euler--Maxwell equations}\label{app:1+2-full}

In section~\ref{sec:1+2-split} we split up the Euler-Maxwell system of equations after considering its ideal MHD limit. Here, we provide for thoroughness the 1+2 decomposition of the full system (no approximations made).

Let us start with Euler's equation in the form~\eqref{eqn:Euler's-full}. Its 1+2 decomposition leads to a scalar (projecting along $n^{a}$)
\begin{eqnarray}
    (\rho+P+\Pi)\mathcal{A}&=&-P'-(\dot{Q}-\alpha^{c}{Q}_{c})-\Theta Q-\Pi'-\frac{3}{2}\Pi\tilde{\Theta}-\tilde{D}^{b}\Pi_{b}+2n'^{b}\Pi_{b}\nonumber\\
    &&+\tilde{\sigma}^{ab}\Pi_{ab}-\Pi^{b} \mathcal{A}_{b}+\mu\epsilon+\epsilon_{bc}j^{b}\mathcal{B}^{c}
    \label{Euler-scalar-full}
\end{eqnarray}
and a vector equation (projection orthogonal to $n^{a}$)
\begin{eqnarray}
(\rho+P+\frac{1}{2}\Pi)\mathcal{A}_{a}&=&-\tilde{D}_{a}P-Q\mathcal{\alpha}_{a}-\dot{Q}_{\bar{a}}-\frac{3}{2}\Theta Q_{a}-\Sigma_{ab}Q^{b}-2\Omega\epsilon_{ab}Q^{b}+3Q\epsilon_{ab}\Omega^{b}\nonumber\\
&&+\frac{1}{2}\tilde{D}_{a}\Pi+\frac{1}{2}\Pi n'_{a}-\Pi'_{a}-\frac{1}{2}\tilde{\Theta}\Pi_{a}-\tilde{D}^{b}\Pi_{ab}+\mathcal{A}\Pi_{a}+\mathcal{A}^{b}\Pi_{ab}\nonumber\\
&&+(\tilde{\omega}_{ab}+\tilde{\sigma}_{ab})\Pi^{b}+\mu\epsilon_{a}-j\epsilon_{ac}\mathcal{B}^{c}+\mathcal{B}\epsilon_{ab}j^{b}\,,
\end{eqnarray}
where we have taken into account that $\Sigma=-\frac{1}{3}\Theta$ and $\Sigma_{a}=-\epsilon_{ac}\Omega^{c}$ (see the previous section). Both 1+2 components of the various quantities as well as the 2-D fluid dynamics fields ($\tilde{\Theta}$, $\tilde{\omega}_{ab}$ and $\tilde{\sigma}_{ab}$) are present in the above relations. It is worth focusing our attention on the last term in the right-hand side of~\eqref{Euler-scalar-full}, namely $\epsilon_{bc}j^{b}\mathcal{B}^{c}$, which vanishes. This happens because $j^{a}n_{a}=0$ and $\mathcal{B}^{a}n_{a}=0$. It is thus clear that the same relation holds for any two vectors which lie on the 2-surface normal to $n^{a}$. The meaning of expression $\epsilon_{bc}j^{b}\mathcal{B}^{c}=0$ is that the vector product of two vectors is not defined in two dimensional space. In our problem, the aforementioned expression implies that there are no forces of magnetic origin affecting the motion along the direction $n^{a}$ of the magnetic field lines.

Regarding Maxwell's equations, their 1+2 split leads to the following components
\begin{eqnarray}
\dot{\epsilon}_{\bar{a}}&=&-\epsilon\alpha_{a}-\frac{1}{2}\Theta\epsilon_{a}-2\epsilon\epsilon_{ac}\Omega^{c}+\Sigma_{ac}\epsilon^{c}+\Omega\epsilon_{ac}\epsilon^{c}-\mathcal{A}\epsilon_{ac}\mathcal{B}^{c}+\mathcal{B}\epsilon_{ac}\mathcal{A}^{c}\nonumber\\
&&-\epsilon_{ac}\mathcal{B}'^{c}-\epsilon_{ac}(\tilde{D}^{c}n_{d})\mathcal{B}^{d}+\epsilon_{ac}\tilde{D}^{c}\mathcal{B}-\mathcal{B}\epsilon_{ac}n'^{c}-j_{a},
\end{eqnarray}
and
\begin{equation}
    \dot{\epsilon}=\epsilon^{a}\alpha_{a}-\Theta\epsilon-2\tilde{\omega}\mathcal{B}+\epsilon_{ac}\tilde{D}^{a}\mathcal{B}^{c}-j
\end{equation}
for the electric field propagation equation as well as
\begin{eqnarray}
\dot{\mathcal{B}}_{\bar{a}}&=&-\mathcal{B}\alpha_{a}-\frac{1}{2}\Theta \mathcal{B}_{a}-2\mathcal{B}\epsilon_{ac}\Omega^{c}+\Sigma_{ac}\mathcal{B}^{c}+\Omega\epsilon_{ac}\mathcal{B}^{c}+\mathcal{A}\epsilon_{ac}\epsilon^{c}-\epsilon\epsilon_{ac}\mathcal{A}^{c}\nonumber\\
&&+\epsilon_{ac}\epsilon'^{c}+\epsilon_{ac}(\tilde{D}^{c}n_{d})\epsilon^{d}-\epsilon_{ac}\tilde{D}^{c}\epsilon+\epsilon\epsilon_{ac}n'^{c}
\end{eqnarray}
and
\begin{equation}
    \dot{\mathcal{B}}=\mathcal{B}^{a}\alpha_{a}-\Theta \mathcal{B}+2\tilde{\omega}\epsilon-\epsilon_{ac}\tilde{D}^{a}\epsilon^{c}
\end{equation}
for the magnetic field propagation equation. Concerning the scalar relations representing Gauss' law for the electric and the magnetic field, their individual terms split leading to
\begin{equation}
    \tilde{D}^{a}\epsilon_{a}+\tilde{\Theta}\epsilon+\epsilon'+n^{a}\epsilon'_{a}+2(\Omega\mathcal{B}+\Omega^{a}\mathcal{B}_{a})=\mu
\end{equation}
and
\begin{equation} 
    \tilde{D}^{a}\mathcal{B}_{a}+\tilde{\Theta}\mathcal{B}+\mathcal{B}'-n'^{a}\mathcal{B}_{a}-2(\Omega \epsilon+\Omega^{a}\epsilon_{a})=0
\end{equation}
respectively. We observe that the full 1+2 decomposed equations are generally more complicated than their original (non decomposed) counterparts. The usefulness of the split in components becomes evident only when specific geometric or physical properties of the problem in hand are taken into account, or even under certain simplifying assumptions reflecting such properties.

\section{Equivalence of the Euler--Maxwell system under the definitions $B^{a}=\mathcal{B}n^{a}$ and $B^{a}=\mathcal{B}k^{a}$ ($n^{a}k_{a}=0$)} \label{sec:Equiv-system}

Let us verify that if we had defined $n^{a}$ to be perpendicular to the magnetic field, namely $B^{a}\equiv \mathcal{B}^{a}=\mathcal{B}k^{a}$ ($k^{a}n_{a}=0$ and $k^{a}k_{a}=1$), we would have arrived at an equivalent system of equations for the magnetised fluid. In particular, we will focus our attention on the vector equations, namely Euler's equations of motion and Faraday's law. Besides, pointing out the equivalence of the scalar equations is a trivial procedure.

First of all, consider Euler's equation in the form of~\eqref{Euler-ideal-MHD-2}. On projecting the latter along $n^{a}$ and setting $B^{a}=\mathcal{B}k^{a}$ (so that $B_{a}n^{a}=0$) we arrive at
\begin{equation}
    (\rho+P+\mathcal{B}^{2})\mathcal{A}=-P'-\mathcal{B}\mathcal{B}'+\mathcal{B}^{2}(k^{c}{\rm D}_{c}k_{a})n^{a}\,,
    \label{Euler-b-par-to-k-1}
\end{equation}
where $-\mathcal{B}\mathcal{B}'$ and $\mathcal{B}^{2}(k^{c}{\rm D}_{c}k_{a})n^{a}$ correspond to the magnetic pressure and tension components of the Lorentz force. Note that $k^{c}{\rm D}_{c}k_{a}$ represents a vector orthogonal to $k^{a}$, namely $n^{a}$. Therefore, the equation in question transforms into
\begin{equation}
    (\rho+P+\mathcal{B}^{2})\mathcal{A}=-P'-\mathcal{B}\mathcal{B}'+\mathcal{B}^{2}\,,
    \label{Euler-b-par-to-k-2}
\end{equation}
which is the equivalent of~(\ref{Euler-ideal-MHD}b). Subsequently, projecting~\eqref{Euler-ideal-MHD-2} orthogonal to $n^{a}$ and setting $B^{a}=\mathcal{B}k^{a}$ as well as $\mathcal{A}_{a}=\mathcal{A}^{*}k_{a}$, we arrive at
\begin{equation}
    (\rho+P+\mathcal{B}^{2})\mathcal{A}^{*}k_{a}=-\tilde{\rm D}_{a}P+\mathcal{A}^{*}\mathcal{B}^{2}k_{a}-\frac{1}{2}\tilde{\rm D}_{a}\mathcal{B}^{2}+\frac{1}{2}(k^{c}{\rm D}_{c}\mathcal{B}^{2})k_{a}\,.
        \label{Euler-b-par-to-k-3}
\end{equation}
Note that $k^{c}{\rm D}_{c}\mathcal{B}^{2}$ represents the norm of the gradient $\tilde{\rm D}_{a}\mathcal{B}^{2}$ and $k_{a}$ its direction, so that $\frac{1}{2}(k^{c}{\rm D}_{c}\mathcal{B}^{2})k_{a}=\frac{1}{2}\tilde{\rm D}_{a}\mathcal{B}^{2}$. As a consequence, our equation finally transforms into
\begin{equation}
    (\rho+P)\mathcal{A}_{a}=-\tilde{{\rm D}}_{a}P\,,
    \label{Euler-b-par-to-k-4}
\end{equation}
which is the equivalent of~(\ref{Euler-ideal-MHD}a) and, as expected, does not include any forces of magnetic origin (no magnetic forces act along the direction of the total magnetic field). In what follows we consider Faraday's and Gauss' law (for the magnetic field),~\eqref{MHD-B-prop} and~(\ref{MHD-constraints}b) respectively. Projecting the former perdendicular to $n^{a}$, subsequently along $k^{a}$, and setting $B^{a}=\mathcal{B}k^{a}$, Faraday's law reads
\begin{equation}
    \dot{\mathcal{B}}=-\frac{1}{2}\Theta \mathcal{B}+\mathcal{B}\Sigma_{ac}k^{a}k^{c}\,.
    \label{Faraday-B-par-to-k-1}
    \end{equation}
Making use of~\eqref{sigma-1+2} we can determine the last term in the right-hand side of the above as
\begin{equation}
    \Sigma_{ac}k^{a}k^{c}=\sigma_{ac}k^{a}k^{c}+\frac{1}{2}\Sigma=-\frac{1}{2}\Theta\,,
\end{equation}
where we have taken into account that $\sigma_{ac}k^{a}k^{c}\equiv {\rm D}_{\langle a}u_{c\rangle}=-\Theta/3$ (${\rm D}_{(a}u_{c)}=0$) and $\Sigma=-\Theta/3$. Hence, equation~\eqref{Faraday-B-par-to-k-1} finally becomes
\begin{equation}
    \dot{\mathcal{B}}=-\Theta \mathcal{B}\,,
\end{equation}
namely eq.~(\ref{eqn:dot-beta+alpha-constr}a), the relation which has led us to the evolution formula for the magnetic field of a highly conducting fluid.\\

\textbf{Acknowledgements:} The present work was supported by the Hellenic Foundation for Research and Innovation (H.F.R.I.), under the ‘First Call for H.F.R.I. Research Projects to support Faculty members and Researchers and the procurement of high-cost research equipment grant’ (Project No. 789). PM also acknowledges support by Leventis Foundation as well as the Foundation for Education and European Culture.\\

\end{document}